\documentclass[graybox, envcountchap]{svmult}

\usepackage{mathptmx}
\usepackage{amsmath}
\usepackage{amssymb}
\usepackage{helvet}
\usepackage{courier}
\usepackage[bottom]{footmisc}
\usepackage{xcolor}
\usepackage{graphicx}
\usepackage{array}
\usepackage{soul}
\newcolumntype{L}[1]{>{\raggedright\arraybackslash}p{#1}}
\usepackage{hyperref}
\hypersetup{colorlinks=true,urlcolor=blue,citecolor=blue,linkcolor=blue}
\usepackage{float}

\begin{document}

\title*{Foundation Models for Astrophysics}
\titlerunning{Foundation Models for Astrophysics}
\author{Xiaosheng Zhao\inst{1} \and Yuan-Sen Ting\inst{2,3,4}}
\authorrunning{X. Zhao \& Y.-S. Ting}
\institute{Xiaosheng Zhao \at Department of Physics \& Astronomy, The Johns Hopkins University, Baltimore, MD 21218, USA, \email{xzhao113@jh.edu}
\and Yuan-Sen Ting \at Department of Astronomy, The Ohio State University, 140 West 18th Avenue, Columbus, OH 43210, USA
\at Center for Cosmology and AstroParticle Physics (CCAPP), The Ohio State University, Columbus, OH 43210, USA
\at Max-Planck-Institut f\"ur Astronomie, K\"onigstuhl 17, D-69117 Heidelberg, Germany, \email{ting.74@osu.edu}}
\maketitle

\abstract{Foundation models are high-capacity networks pretrained once on broad data and then reused across many tasks. This chapter introduces them through the idea of a \emph{transferable representation}, the internal description a network forms during training, which, rather than the fitted task, is what carries over to new problems. We develop the idea from first principles for an astronomical reader, starting from why a representation matters and what makes one useful, and then surveying the architectures, self-supervised objectives, scaling, adaptation, and cross-modal learning that produce one. A theme throughout is the distinction between these methods and the goal they serve. The presence of a transformer, a self-supervised objective, and large-scale pretraining does not by itself make a model a foundation model, since the defining property is that the learned representation transfers, as tested by its ability to work on new tasks with little or no task-specific training data (few-shot and zero-shot learning). We then consider astronomy, where data are abundant but labels are scarce and simulations often stand in for ground truth. Here we offer a cautious reading of the current literature, in which many models adopt the architecture of foundation models while clear demonstrations of transfer across instruments, populations, and tasks remain comparatively rare. This is to be expected, since robust transfer beyond language is still uncommon even in vision and the wider physical sciences, and whether further scaling or a different account of representation will close the gap remains an open question. We close by placing the goal within the broader aim of machine intelligence and outlining the evidence that would mark real progress.}

\section{Representations that are transferable}

\begin{figure}[t]
\centering
\includegraphics[width=\textwidth]{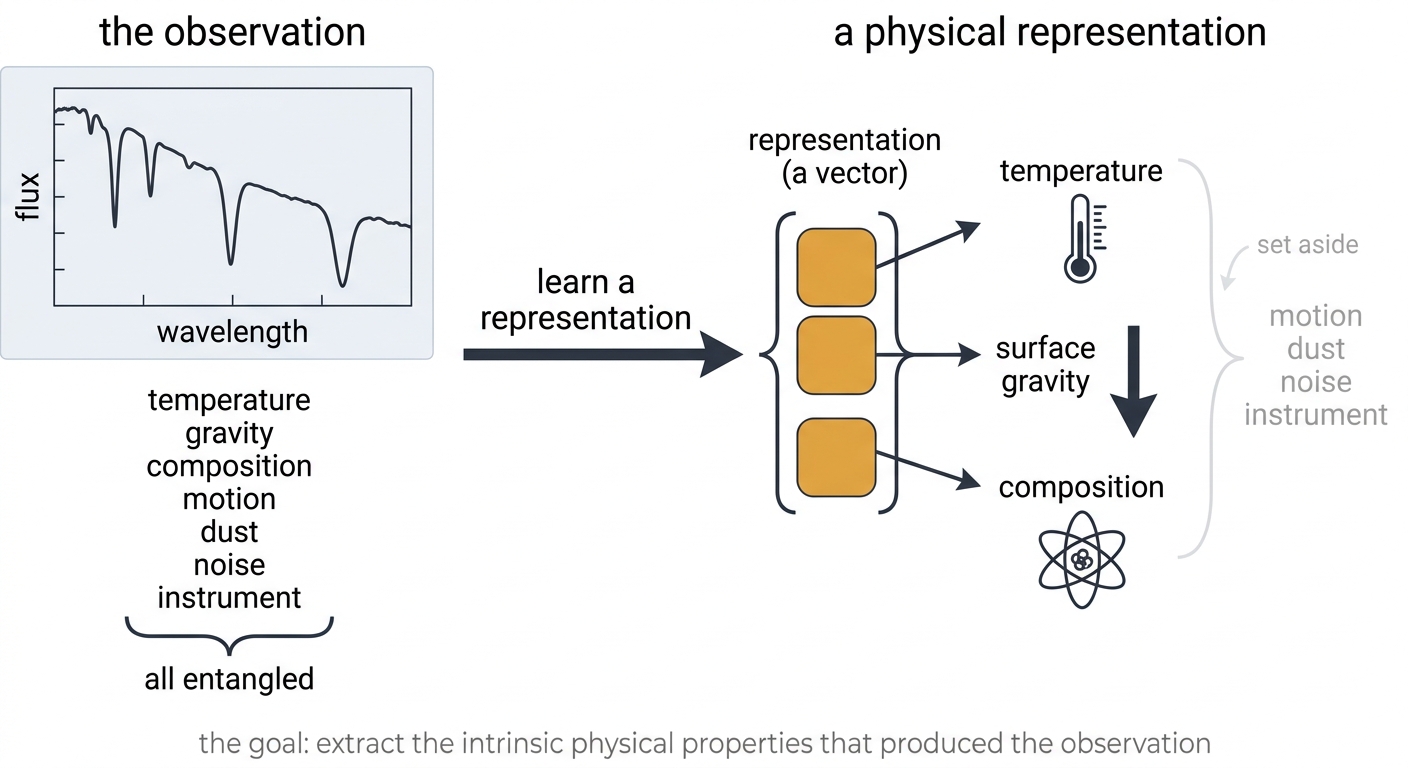}
\caption{The goal of a representation. An observation entangles the intrinsic physical properties with motion, noise, and the instrument, and the aim is to learn a \emph{representation}, a vector whose coordinates map to those intrinsic properties, here temperature, surface gravity, and composition, with the extrinsic and nuisance effects set aside. A representation organised this way, with the physics pulled out and the rest discarded, is one that can transfer, staying useful when the instrument, the population, or the task changes.}
\label{fig:goal}
\end{figure}

To many astronomers, deep learning looks like a flexible form of regression or classification. A neural network is a function with many adjustable parameters, trained to turn an input into an output, and from that angle it is an elaborate cousin of a polynomial fit or a random forest. For a single task with enough labelled examples, that description is fair. What it leaves out is the reason for the recent shift toward \emph{foundation models}, which are trained once on a large and varied body of data and then reused for many tasks, including ones their builders never had in mind. The shift makes sense once we notice that a network does two things while it trains. It fits the task in front of it, and along the way it builds an internal description of the input. That internal description, which we will call a \emph{representation}, is what gets reused, and learning a good representation is a different goal from fitting any single task (Figure~\ref{fig:goal}).

\begin{figure}[t]
\centering
\includegraphics[width=\textwidth]{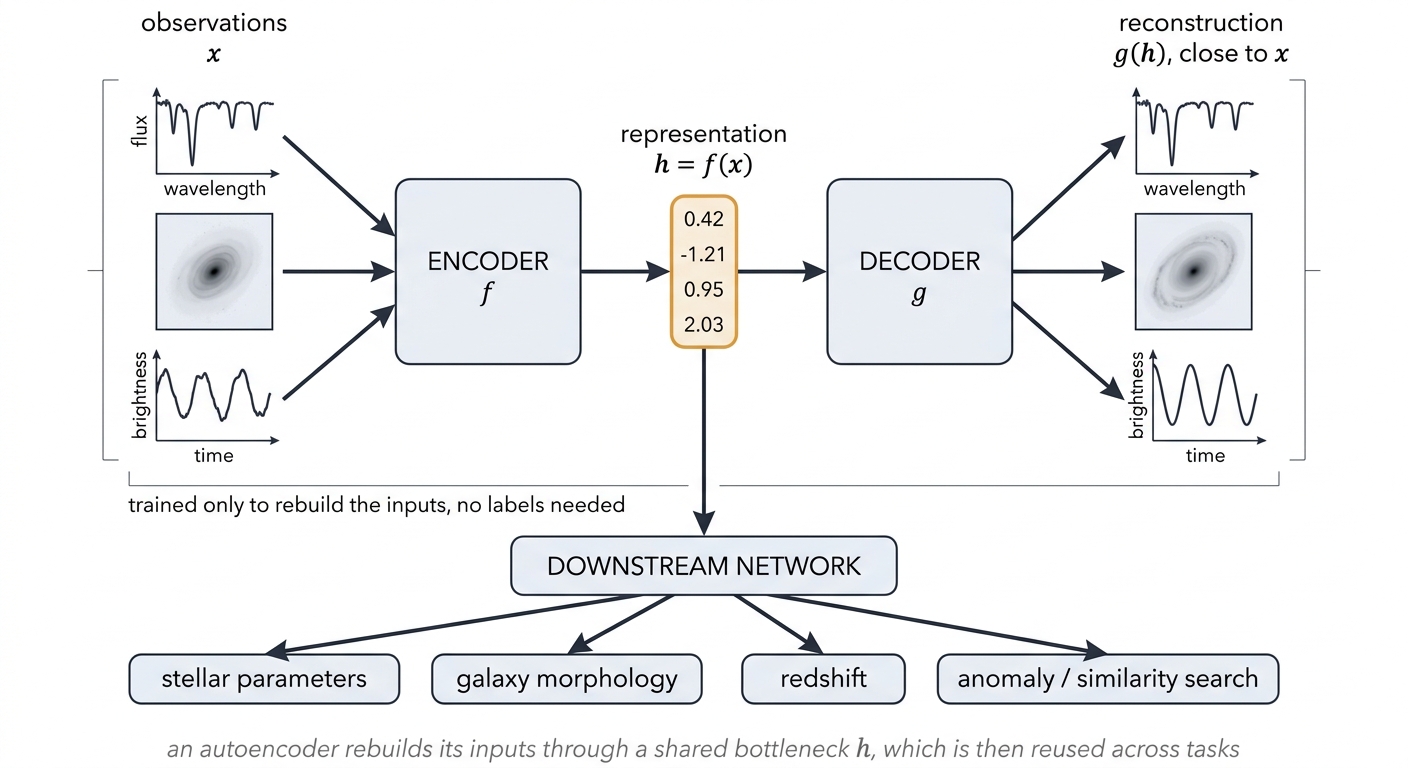}
\caption{Learning a shared representation, and reusing it. An \emph{autoencoder} is one common way to obtain a representation without labels. An encoder $f$ compresses observations $\mathbf{x}$, here a spectrum, an image, and a light curve, through a narrow bottleneck $\mathbf{h}=f(\mathbf{x})$, while a decoder $g$ is trained only to rebuild them, so that $g(\mathbf{h})\approx\mathbf{x}$. The bottleneck $\mathbf{h}$ is the reusable product, fed to a downstream network for tasks such as estimating stellar parameters, classifying morphology, or finding anomalies. This is not the only route to a representation, but a representative one.}
\label{fig:foundation_model}
\end{figure}

The idea behind a representation is already familiar in astronomy, even if the name is not. An observation arrives as raw numbers set by the instrument, the flux in each wavelength bin, the counts in each pixel, or the brightness at each epoch. These are seldom the quantities we reason with, so we routinely recast them as more revealing descriptions such as colours, line indices, periodograms, or fitted physical parameters, each of which keeps some of the information and sets the rest aside. A representation is this same step taken automatically. Writing the raw observation as $\mathbf{x}$, the network applies a transformation $\mathbf{h}=f(\mathbf{x})$. The map $f$ is called the \emph{encoder}, and the vector $\mathbf{h}$ it produces, the \emph{representation} or \emph{embedding}, is a new description of the observation, learned from the data rather than chosen by hand~\cite{bengio2013representation} (Figure~\ref{fig:foundation_model}).

Descriptions are not equally useful, and that difference is what makes a representation worth having. A good one brings the quantities we care about within easy reach and pushes the rest into the background. In a stellar spectrum the raw flux blends temperature, surface gravity, chemical composition, motion along the line of sight, dust along the path, photon noise, and quirks of the spectrograph. A useful representation untangles this enough that the physical quantities can be read off by a simple downstream step, while detector noise and calibration drifts no longer dominate. It cannot conjure information the observation never held. Instead, it makes the present information easier to use.

Reducing data to a few informative numbers is not new. Principal-component analysis~\cite{hotelling1933analysis} and factor analysis~\cite{bartholomew2011latent} both do it with linear algebra, looking for a small set of shared patterns and describing each observation $\mathbf{x}$ as a weighted combination of them, $\mathbf{x} \approx \mathbf{W}\mathbf{h}$, where the columns of the matrix $\mathbf{W}$ are the shared patterns and the vector $\mathbf{h}$ holds the few coefficients that weight them, far fewer numbers than the original $\mathbf{x}$. Principal-component analysis chooses the patterns that account for the most variance across the dataset, while factor analysis treats them as hidden factors that generate the data up to some per-measurement noise~\cite{christopher2006pattern}. Because the patterns are combined linearly, they capture the dominant correlations in the data but not the physical factors that generated it, which seldom lie along straight directions in the raw numbers.

Neural networks lift that restriction. Their internal features are nonlinear and are shaped by the training task itself, so they can follow curved structure that a linear method cannot, and they build the features in layers, with simple local patterns early and more elaborate combinations later, an idea developed over decades from early neuron models and the perceptron to back-propagation and convolutional networks~\cite{mcculloch1943logical,rosenblatt1958perceptron,rumelhart1986learning,fukushima1980neocognitron,lecun1998gradient}. A downstream predictor then works from these learned features rather than from the raw pixels~\cite{krizhevsky2012imagenet,zeiler2014visualizing}, which is what is meant by saying that a deep network learns a representation while it learns its task. What is surprising is that this turned out to be more than a gimmick. Scaled up, these nonlinear models do not merely fit their training data more closely. They appear to capture something more fundamental than the correlations a linear method would find.

\subsection{The language success story}

Whether a representation is any use beyond the data it was trained on is the question a foundation model has to answer, and language is where it has been answered. It helps to be precise about what ``working'' means. Although \emph{generalization} and \emph{transfer} are often used interchangeably, what matters is whether evaluation is \emph{in-domain} (evaluating on data drawn from the same distribution as the training data, or \emph{in-distribution}) or \emph{cross-domain} (evaluating on a different distribution). A model \emph{generalizes} when it does well on new examples drawn from the same population as its training set (in-domain), and it \emph{transfers} when it stays useful after something changes, a different task, a different instrument, or a different population (cross-domain). The cross-domain case is the harder one, and the one foundation models are built around. The recipe is to \emph{pretrain} once on broad data, learning a representation with no single task in mind, and then \emph{adapt} it as the target or the data change~\cite{bommasani2021foundation}, scaling up the older ideas of transfer learning and multi-task learning~\cite{caruana1997multitask,pan2009survey}.

\begin{figure}[t]
\centering
\includegraphics[width=\textwidth]{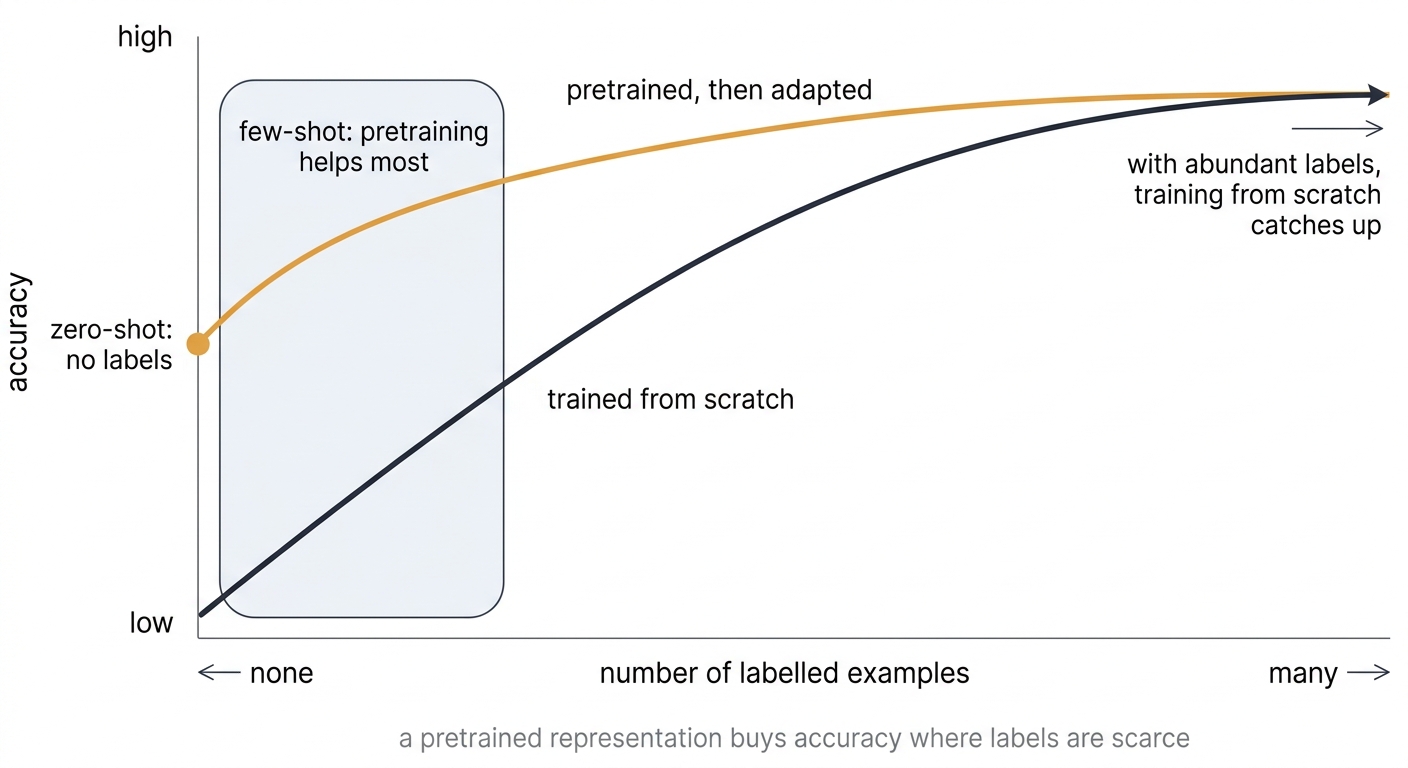}
\caption{Why a transferable representation is worth having. The vertical axis is accuracy on a downstream task, a stand-in for how well that task is done, classification accuracy or an equivalent score for a regression. As the number of labelled examples grows, a model trained from scratch (charcoal) improves from a low base, while a pretrained model that is only adapted (amber) starts far higher and reaches good accuracy from a handful of labels, the \emph{few-shot} regime; with no labels at all it can still work \emph{zero-shot}, the point on the vertical axis. The advantage is largest where labels are scarce and narrows as they grow abundant, where training from scratch eventually catches up.}
\label{fig:label_efficiency}
\end{figure}

The way to test for transfer is to give the model a task it was not trained for and count how much task-specific labelling it still needs. Working with no labels is called \emph{zero-shot} learning, which draws only on what pretraining built. Working with a small handful of examples is \emph{few-shot} learning, the situation astronomy meets whenever a rare class of object has just a few confirmed members (Figure~\ref{fig:label_efficiency}). Large language models clear this bar, performing tasks no one trained them on---zero-shot, or from a few examples placed in the prompt~\cite{brown2020gpt3}.

That success runs deeper than any one task. A single pretrained network can translate, answer questions, summarise, and write code, and probes of its internal activations turn up structure no one built in by hand, including grammatical features and even approximately linear encodings of quantities such as space and time~\cite{cunningham2024saes,gurnee2024spacetime}. Exactly how a network arrives at such structure is only partly understood, and reading it back out has become a research area of its own, \emph{mechanistic interpretability}~\cite{Sharkey2025mechanistic}.

That this structure is learned, rather than the training set simply being memorised, can be seen most directly at small scale, in an effect called \emph{grokking}. Consider a network trained on a sharply defined task, such as adding two integers modulo a prime. Early on it behaves like any overfit model, reproducing the correct output for each integer pair it was shown while doing no better than chance on the pairs held out. If training continues well past the point where the training error has already fallen to zero, the test accuracy, stuck near chance for a long time, can rise abruptly as the network stops memorising and begins to apply the general rule, now succeeding on pairs it has never seen~\cite{power2022grokking,nanda2023progress}. What underlies this behavioral shift is visible in the weights: following the weights across this transition shows a general-purpose circuit taking shape while the memorised solution is pruned away~\cite{nanda2023progress,varma2023circuit}. Fitting the data and learning a rule that transfers are, in this sense, two different internal states, and training can carry a network from the first to the second. A loosely analogous delayed appearance recurs at much larger scale in language models, where a capability can appear fairly abruptly once the model passes a certain size, an \emph{emergent ability}~\cite{wei2022emergent}.

\subsection{Why astronomy needs transfer}

While language shows that transfer can be achieved, astronomy is a domain where it is also needed. Two features of its data make a transferable representation worth the effort, and they reinforce each other.

The first is that labels are scarce. Modern surveys are enormous. The Vera C.\ Rubin Observatory will catalogue tens of billions of sources~\cite{Ivezic2019} and \textit{Euclid} will image more than a billion galaxies~\cite{Euclid2025}, yet trustworthy labels exist for only a small and often biased fraction of them. The asymmetry is built into how labels are made. Some are assigned by hand, and the citizen-science morphologies of Galaxy Zoo, for example, cover only a portion of the galaxies already imaged, with human effort unable to keep pace with the surveys~\cite{Masters2019galaxyzoo,Walmsley2022jointschedule}.

Others are physical parameters that come from a more expensive measurement, or from fitting a physical model to such a measurement. A common pattern is label transfer, where a smaller set of objects has higher-quality spectra or parameters inferred from detailed spectral modelling, and those labels are then used to train a model for a much larger set of cheaper observations, such as lower-resolution spectra. This scales the label set, but it also transfers the limitations of the reference instrument, the selection of the overlap sample, and the physical model used to assign the original labels~\cite{Kurucz1993,garciaperez2016aspcap}. A representation learned from the plentiful unlabelled observations can lower how many such reference labels are needed, and can test whether the same information carries from one instrument to another.

The second is the gap between simulations and observations. Physics-based forward models (simulations) can generate many labelled examples in regimes where real labels barely exist, so they are leaned on heavily; in cosmology, where we observe only one universe, high-fidelity simulations often stand in as ground truth for training and validation. But synthetic data are not the sky. A model trained on simulations has to cross over to real instruments, where the input physics, the noise, the instrumental response, the calibration, and the underlying population may all differ from what the simulator assumed. Closing this synthetic--observed gap is itself a transfer problem~\cite{OBriain2021cyclestarnet}, so a representation that survives a change of domain is worth more than one tuned to a single simulated set.

\section{Learning transferable representations}

Knowing why a transferable representation is worth having does not say how to build one, and there is no recipe that guarantees it. What we can do is name the properties a good representation tends to have, and then the design choices that, in practice, help produce them.

If a representation is the thing we want to reuse, it helps to say what a reusable one looks like. There is no single number that grades representation quality, since what counts as good depends on the task, but a few properties recur. The central one is that the quantities relevant across many tasks stay accessible while irrelevant variation is held down. The two kinds of variation are worth naming. A \emph{factor of variation} is a quantity whose change produces a systematic change in the observation. In astronomy, this may be physical, such as temperature or redshift, or observational, such as seeing or signal-to-noise. A \emph{nuisance} is a factor we want the representation to ignore for the task in hand, even if it matters for a different one.

\begin{figure}[t]
\centering
\includegraphics[width=\textwidth]{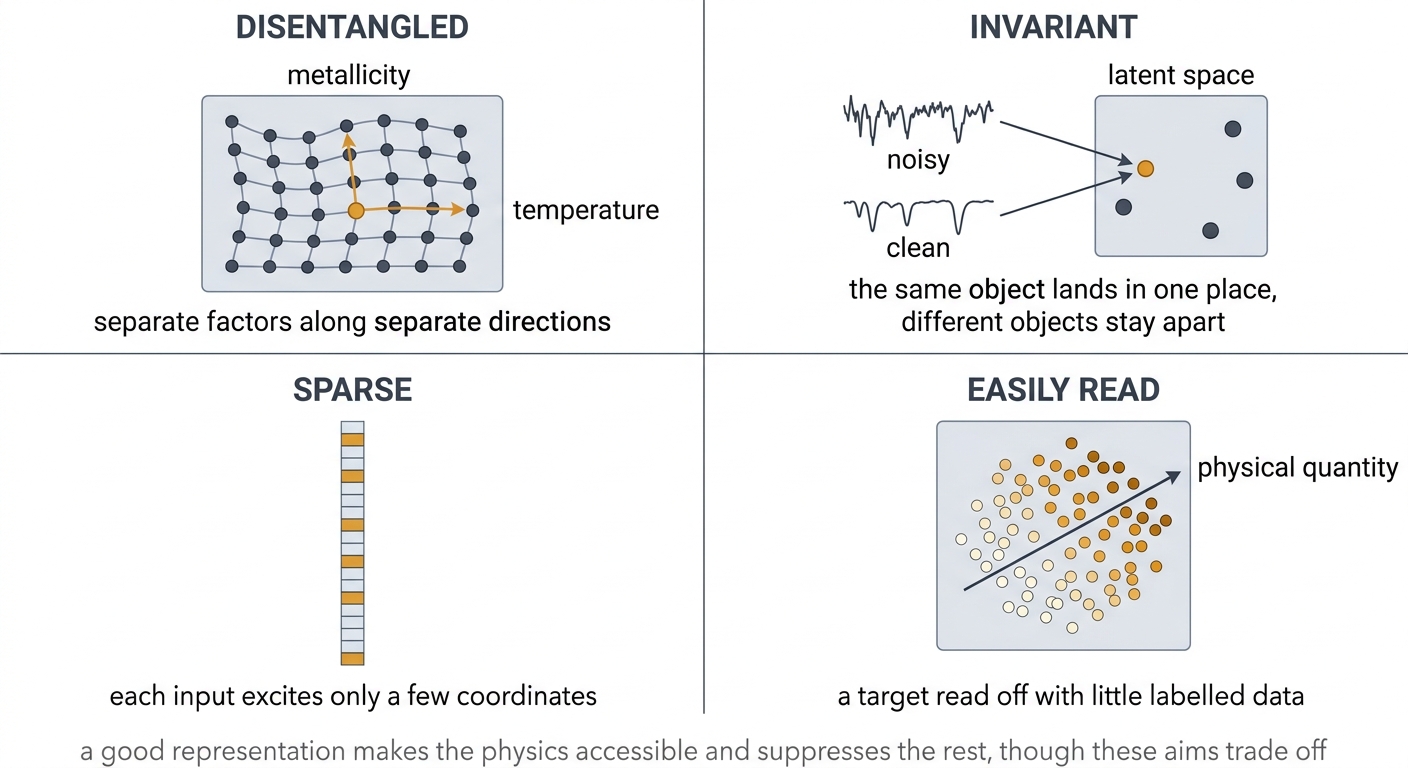}
\caption{Properties that tend to make a representation good. It is \emph{disentangled} when separate physical factors lie along separate directions; \emph{invariant} when the same object maps to the same point whatever the noise or instrument; \emph{sparse} when each input excites only a few coordinates; and \emph{easily read} when a target follows from the representation with little labelled data. These aims do not reinforce one another, since suppressing everything about the instrument can also discard information a later task needs.}
\label{fig:properties}
\end{figure}

The property that matters most in practice is the plainest one: coordinates from which a target can be predicted using only a little labelled data---\emph{easily read}, in practice measured by how well a linear probe does with few labels. This is what transfer means operationally, and it is what the rest of this review will keep returning to. Alongside it sits \emph{invariance} to chosen nuisances, so that the same object maps to the same place whatever the noise or the instrument. Two further ideals are worth naming as well, since they shape how people think about the problem even where they are only partly attained. A recurring ideal is \emph{disentanglement}, in which the separate factors of variation end up along separate directions of the representation, so that changing one physical quantity moves the embedding one way and leaves the others undisturbed~\cite{bengio2013representation} (Figure~\ref{fig:properties}). \emph{Sparsity}, where each input excites only a few coordinates, can make the active factors easier to read off~\cite{olshausen1996emergence}. Neither comes for free from the objectives used at scale: the representations of large language models are famously neither disentangled nor sparse, which is why recovering sparse, interpretable directions from them is itself an active research programme~\cite{cunningham2024saes}. The sharpest tension is between invariance and sufficiency. Discarding everything about the instrument may aid transfer between surveys, yet that same instrumental information may be just what is needed to model the noise or the selection function. A representation therefore has to keep enough for its intended uses while suppressing only what is known to be irrelevant.

Picture what this would mean for a stellar spectrum. An ideal embedding would let one recover the parameters of spectral synthesis: effective temperature, surface gravity, metallicity and individual elemental abundances, rotation, and radial velocity. At the same time, noise and calibration residuals should not drive the distances between stars in the embedding. It would then act as a partial inverse of spectral synthesis, mapping an observed spectrum back toward the quantities that could have produced it. This does not happen on its own. It may take training on spectra from several instruments, augmenting the data with realistic noise and calibration shifts, or adding objectives that separate physical factors from observational ones. That list is not arbitrary: paired views of the same object, whether two augmentations or two instruments, are close to the conditions under which the underlying factors are provably recoverable at all---the question of identifiability~\cite{hyvarinen1999nonlinear, seflsupervised2021, Daunhawer2023}. Whether such separation is attainable in astrophysical survey data is an open question.

Much of the toolkit that follows, the transformer, self-supervised pretraining, scaling, and cross-modal alignment, was carried forward on the strength of language and vision models and then borrowed by other fields, astronomy among them, building on a long line of work on representing words and their context~\cite{jelinek1997statistical,mikolov2010recurrent,hochreiter1997lstm,mikolov2013word2vec,sutskever2014seq2seq,bahdanau2015attention,vaswani2017attention,peters2018elmo,devlin2018bert,chen2020simclr,wang2024llmsurvey}. That success shows these choices can in practice produce representations that transfer, which is the property we actually need, so they are worth taking seriously as guidelines. They are not guarantees, though, and the goal should not be confused with the toolkit. Assembling the same parts, a large transformer trained with a self-supervised objective on a big dataset, does not by itself make a foundation model; what makes one is that the representation actually transfers. The parts are only the means to that end, a distinction easy to lose in astronomy as much as anywhere. With that in mind, we take the main choices in turn. The vocabulary used in what follows is collected in Table~\ref{tab:glossary}.

\begin{table}[H]
\caption{A short glossary of the foundation-model terminology used in this chapter, with rough astrophysical counterparts.}
\label{tab:glossary}
\renewcommand{\arraystretch}{1.08}
\footnotesize
\begin{tabular}{@{}L{2.7cm}L{5.1cm}L{3.3cm}@{}}
\hline\noalign{\smallskip}
\textbf{Term} & \textbf{Meaning} & \textbf{Astrophysical analogue} \\
\noalign{\smallskip}\svhline\noalign{\smallskip}
Foundation model & A high-capacity network pretrained once on broad data and adapted to many downstream tasks & A single pretrained encoder reused across surveys and tasks \\
Representation / embedding & The internal learned vector encoding an input; ``embedding'' the specific vector, ``representation'' the general idea & A compact code for a spectrum or image from which physical parameters are read \\
Backbone & The primary pretrained neural network that extracts features/representations from raw inputs, which is then kept frozen or adapted & The core neural network model that processes spectra or images, whose outputs are fed to simpler predictors \\
Generalization / transfer & Doing well on new examples from the training distribution (in-domain); staying useful after a change of task, instrument, or population (cross-domain) & Working on new objects from the same survey, versus still working on a different survey or instrument \\
Zero-shot / few-shot & Using no, or only a few, task-specific labelled examples & Recognizing a rare class of object from a handful of confirmed members, or from none \\
Inductive bias & Built-in assumptions about data structure that improve data efficiency (learning from fewer examples) & Translation invariance for images; physical symmetries such as rotation invariance \\
Attention / transformer & An architecture in which each element becomes a data-weighted blend of the others, linking distant positions directly & Combining a single element's absorption lines spread across wavelength \\
Self-supervised pretraining & Learning from targets built from the unlabelled data itself, through an auxiliary or surrogate pretext task & Pretraining on unlabelled spectra or light curves before any reference labels are used \\
Contrastive / joint-embedding & Pulling matched views of one object together in the representation and (optionally) pushing unmatched ones apart & Aligning two instruments' spectra of one source, or an image and a spectrum of one galaxy \\
Neural scaling laws & The power-law fall of test loss as model size, data, and compute grow in balance & Measured for stellar-spectra emulators across model, data, and training size \\
Fine-tuning & Adapting a pretrained model to a task by updating all weights, a low-rank correction (LoRA), or only a linear output layer (linear probe) & Adapting a pretrained model to a new survey from a small labelled set \\
Modality / modality gap & A kind of measurement; the modality gap is what remains when two measurement types are mapped into the same representation but do not line up & Image, spectrum, light curve, catalogue; the gap when their encoders do not align \\
\noalign{\smallskip}\hline
\end{tabular}
\end{table}

\subsection{Architectures and inductive biases}

The first of these choices is the \emph{architecture}, the pattern of connections that decides how parts of the input are allowed to interact. A transferable representation needs no particular architecture, nor even self-supervised training; any supervised network builds intermediate features, and some of them survive a change of task or dataset~\cite{yosinski2014transferable,kornblith2019better}. In spectroscopy, for instance, even a plain fully connected network, the basic kind in which every input feeds every unit, can transfer from one spectral resolution to another with only a small fine-tuning set~\cite{Zhao2026generalize}.

Architecture matters because it makes some functions easy to learn and others hard. A wide enough network with a single hidden layer can in principle approximate any continuous function~\cite{cybenko1989approximation,hornik1991approximation}, but that guarantee says nothing about how much data or training reaching it would take. What makes an architecture practical is its set of built-in assumptions, its \emph{inductive biases}. Each family is tuned to a kind of data, with spatial locality for images, an ordering for time series, relational structure for catalogues, and long-range dependence for sequences. A convolutional network (CNN), for example, assumes that the same local pattern is worth looking for wherever it appears in an image, much as an astronomer detects sources by matched-filtering an image with a fixed kernel, except that the network learns its filters from the data instead of fixing them in advance. This is what keeps a galaxy a galaxy after it shifts on the detector. A graph network instead assumes the answer should not depend on the arbitrary order in which nodes are listed. This suits catalogues or particle sets with no natural ordering~\cite{battaglia2018relational,bronstein2021geometric}.

\begin{figure}[t]
\centering
\includegraphics[width=\textwidth]{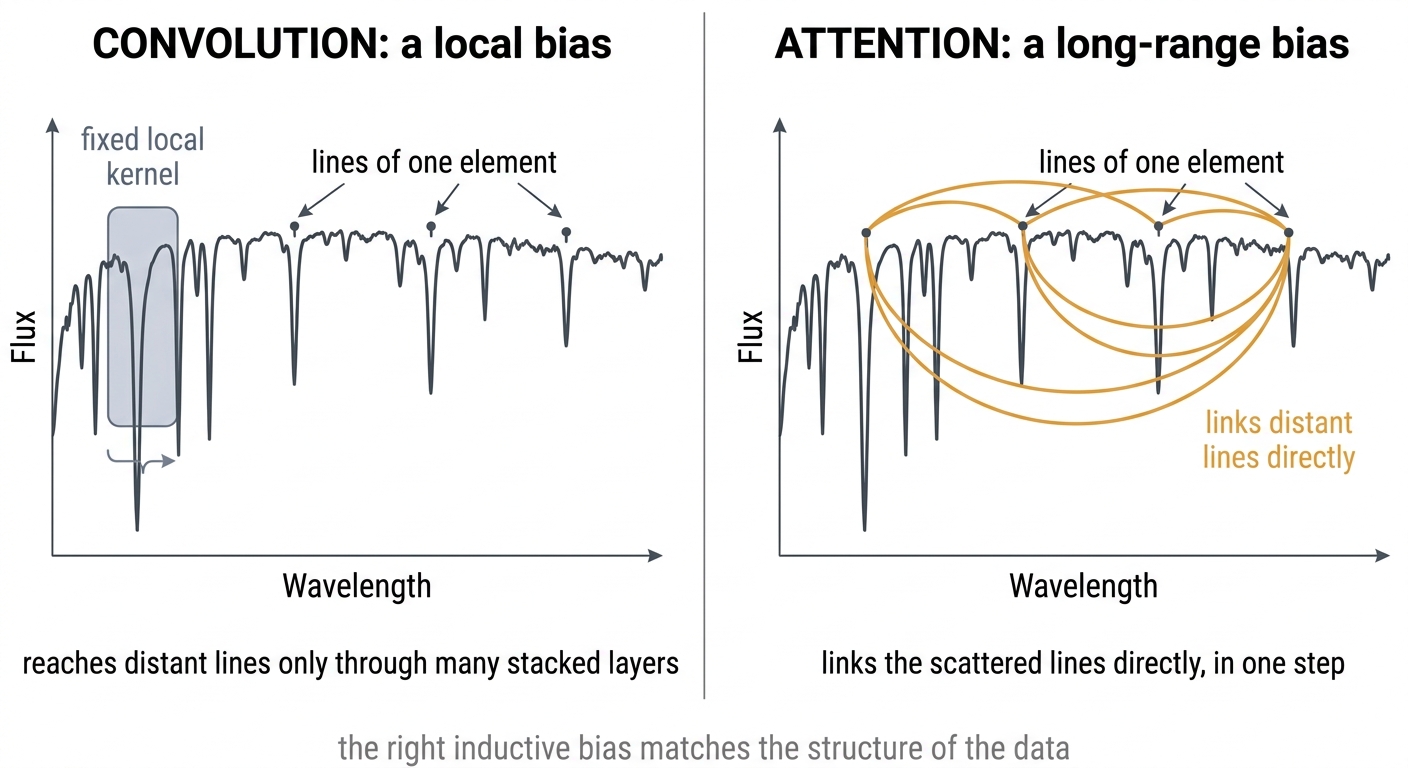}
\caption{Inductive bias, the structure an architecture assumes in its input, shown for a stellar spectrum in which the lines of one element fall scattered across wavelength. A \emph{convolution} carries a local bias, since each kernel sees only neighbouring wavelengths, so it ties distant lines together only indirectly, by stacking many layers. \emph{Attention} carries a long-range bias, linking distant positions directly, in a single step, so an element's scattered lines connect at once. When the bias matches the structure of the data, less of that structure has to be learned from the data alone.}
\label{fig:inductive_bias}
\end{figure}

The transformer, the architecture behind large language models, is built on \emph{attention}, in which each element of the input forms a weighted blend of the others, the weights computed from the data. This suits a spectrum, since the lines of a single element are scattered across wavelength and measuring its abundance means combining all of them, each weighted by how diagnostic it is. A convolution is a poorer match, because its fixed local kernel blends neighbouring wavelengths that often encode different physics, whereas attention can tie an absorption feature directly to every related line however far apart they fall~\cite{rozanski2025scaling} (Figure~\ref{fig:inductive_bias}), and the same mechanism handles time series, whose measurements are often spaced unevenly in time~\cite{ting2026deeplearning}.

 What attention gives up is a built-in sense of order, so position or wavelength has to be supplied alongside the data~\cite{vaswani2017attention,dosovitskiy2021vit}. Its assumptions are lighter than a convolution's, closer to treating the input as an unordered set~\cite{lavie2024inductivebias}. Convolutions, graphs, and transformers differ not just in size but in the structure they make easy to express.

\subsection{Self-supervised pretraining}

The architecture fixes what a network can compute; the training objective decides which of those computations it is rewarded for, so the same network can end up with very different representations depending on what it is asked to predict. In ordinary supervised learning the target comes from outside, as a class label or a measured parameter. In \emph{self-supervised} learning the network instead manufactures its own target from the input, through an auxiliary or surrogate task (often called a \emph{pretext task}). A stretch of a spectrum can be hidden and then predicted from the rest, two observations of one source can be set up as a matching pair, or one part of a light curve can be predicted from another. No human labels are needed, which is what makes the approach attractive when labels are scarce.

\begin{figure}[t]
\centering
\includegraphics[width=\textwidth]{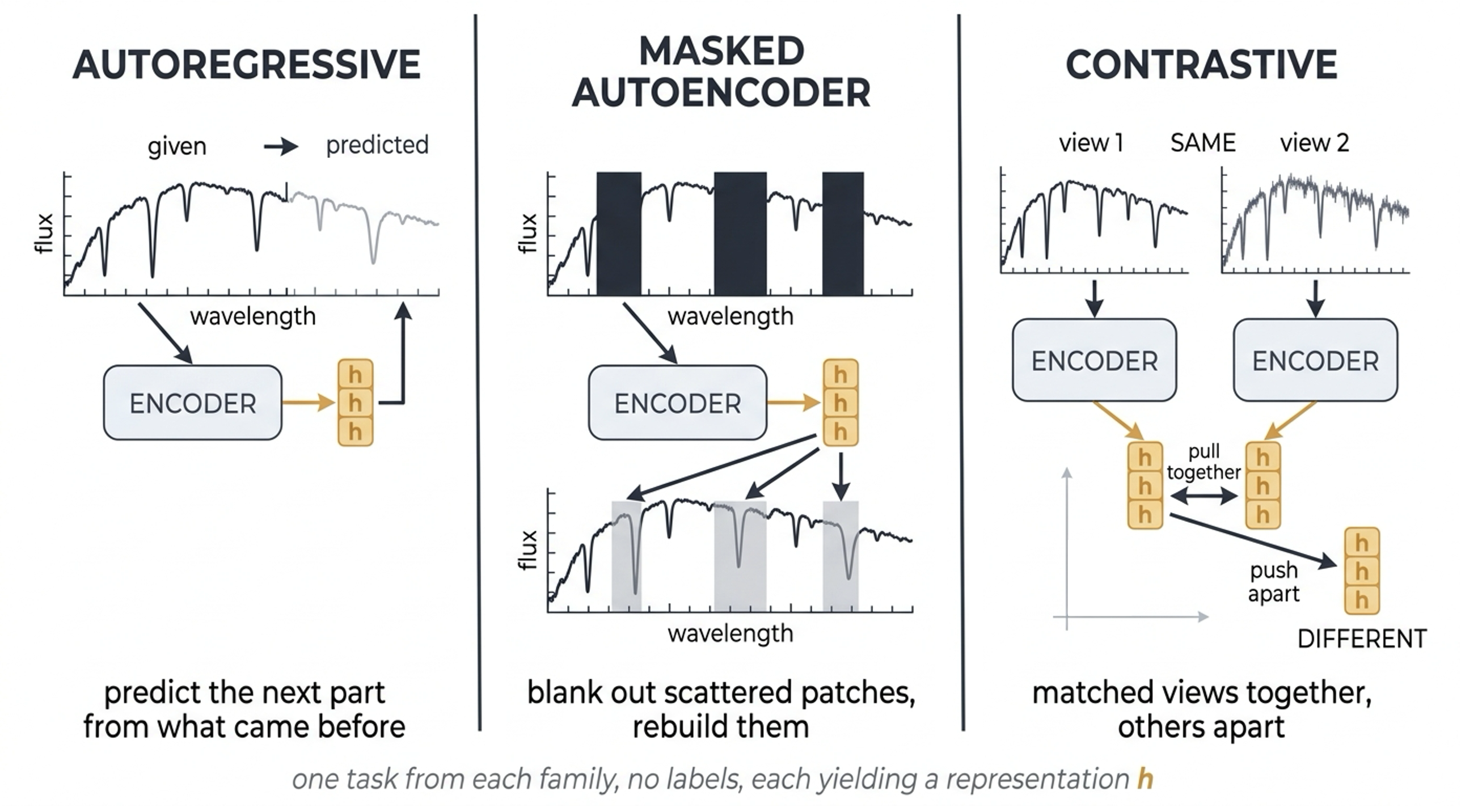}
\caption{One representative self-supervised pretext task from each of the three families, each learning a representation $\mathbf{h}$ from unlabelled spectra without physical labels. \emph{Autoregressive} prediction extends the input, filling in the next part from what came before. \emph{Masked autoencoding} blanks out scattered patches and rebuilds them. \emph{Joint embedding}, shown here with a contrastive loss, pulls two views of one object together while pushing different objects apart. The objective differs, but each yields a representation $\mathbf{h}$ that can be reused downstream.}
\label{fig:ssl_families}
\end{figure}

This self-supervised stage is usually followed by a shorter supervised one once labels are available. Pretraining organizes the data before any particular physical quantity is requested, and it can help even when both stages see the same inputs, because it sets a better starting point~\cite{erhan2010why}; whether it helps in a given case depends on how much of the information the downstream task needs is kept by the self-supervised objective. The objectives in use fall into three broad families (Figure~\ref{fig:ssl_families}).

The first family predicts missing or future parts of the data. Autoregressive language models predict the next \emph{token}, the unit a model reads and predicts, typically a word or part of a word, from the ones before it. In symbols, they maximize the log-likelihood $\sum_t \log p(x_t \mid x_1,\dots,x_{t-1})$ summed over positions $t$, where $p$ is the probability assigned to the token $x_t$ given the earlier tokens $x_1,\dots,x_{t-1}$~\cite{radford2019gpt2,brown2020gpt3}. The spectral analogue would predict the next wavelength bin, though spectra have different noise and a different kind of ordering. These methods are \emph{generative}, meaning that once trained they can produce new samples rather than only encode existing ones.

The second family reconstructs the input through a bottleneck. An \emph{autoencoder} joins an encoder $\mathbf{h}=f(\mathbf{x})$, which compresses the input to a representation, with a decoder $\hat{\mathbf{x}}=g(\mathbf{h})$ that rebuilds it, trained to minimize a reconstruction loss such as $\mathcal{L}_{\mathrm{AE}}=\mathbb{E}_{\mathbf{x}}\lVert \mathbf{x}-g(f(\mathbf{x}))\rVert^2$, where $\lVert\cdot\rVert$ is the Euclidean norm and $\mathbb{E}_{\mathbf{x}}$ the average over inputs, so the loss is the mean squared difference between each input and its reconstruction, while $\mathbf{h}$ is kept smaller than the input. This is the nonlinear counterpart of the linear factorization $\mathbf{x}\approx \mathbf{W}\mathbf{h}$. With linear $f$ and $g$ the best solution is simply the principal-component subspace~\cite{Baldi1989}, whereas nonlinear $f$ and $g$ can wrap a curved data surface onto fewer coordinates. For a stellar spectrum, the aim is that the bottleneck keeps the few numbers that set the line strengths, temperature, gravity, and composition, while the per-pixel noise it cannot predict falls away. Left unconstrained, the code can waste capacity copying irrelevant detail, so variants add constraints or regularization of different kinds, among them contractive, denoising, and sparse autoencoders~\cite{bengio2013representation}.

A close relative is masked modeling, where only the hidden portion has to be rebuilt~\cite{devlin2018bert,he2022mae}. Filling a masked wavelength region forces a model to use how lines vary together, since strong iron lines in one part of a spectrum imply iron lines of correlated strength elsewhere, and in learning to fill the gap the model takes in how temperature, density, and composition shape the spectrum, without any of these being supplied as a label.

The third family is \emph{joint embedding}, which lines up different views of the same thing in the representation. The views might be two augmentations of one image, an image and its caption, or spectra of one star from two instruments. Contrastive learning is one way to do this: it pulls matched views together and pushes unmatched ones apart~\cite{oord2018cpc}, and CLIP applies exactly this to images and text~\cite{radford2021clip}. When the matched views are spectra of one star taken by two instruments, agreement forces the representation to keep what the instruments share, the temperature, gravity, and composition of the star, and to discard what they do not, each instrument's resolution, noise, and calibration, so what survives is the part that transfers from one to the other. The choice of views therefore decides what the model is told to treat as the same object.

Writing $\mathbf{z}_{x}=f(\mathbf{x})$ and $\mathbf{z}_{y}=g(\mathbf{y})$ for the representations of two matched views, and $\mathbf{z}^{-}_{k}$ for unmatched ones, a loss such as information noise-contrastive estimation (InfoNCE) takes the form
\begin{equation}
\mathcal{L} = -\log \frac{\exp\!\big(\mathrm{sim}(\mathbf{z}_{x},\mathbf{z}_{y})/\tau\big)}{\exp\!\big(\mathrm{sim}(\mathbf{z}_{x},\mathbf{z}_{y})/\tau\big) + \sum_{k}\exp\!\big(\mathrm{sim}(\mathbf{z}_{x},\mathbf{z}^{-}_{k})/\tau\big)},
\end{equation}
where $\mathrm{sim}$ is a similarity, usually the cosine, and $\tau$ a temperature that sets how sharply matched and unmatched pairs are separated. The unmatched examples, the \emph{negative pairs}, are what stop the network from collapsing every input onto the same point, but they also make the result sensitive to how the negatives are chosen. Some joint-embedding methods avoid negative pairs. BYOL and DINO, for example, train one version of a network to match the representation produced by another version of the same network, a strategy often called self-distillation (where a network acts as its own teacher, predicting its own outputs under different views to prevent the representation from collapsing into a single constant value)~\cite{grill2020byol,caron2021dino}. The families thus differ in where the loss falls, on the predicted next or missing value, on the reconstructed input, or on the agreement between two views, and each defines ``similar'' differently.

Each of these objectives, on its own, captures only part of what governs a spectrum, and none is given the physical parameters. The hope, though not a guarantee, is that combining them constrains the representation from several directions at once, so that what survives all of them approaches the small set of physical factors that generated the data, the disentangled and transferable description an ideal representation aims at. Whether a given combination reaches that, or only a convenient compression that must still be judged against the scientific differences one needs to keep, is what transfer has to decide.

\subsection{Combining modalities}

Contrastive pretraining aligns two views of the same object, and the views need not be the same kind of data. Astronomy rarely captures an object through a single instrument, and an image and a spectrum of one galaxy are two such views, carrying overlapping but complementary physics. The image records its size, shape, and orientation, while the spectrum records the composition, ionization, and motion of its gas and stars. Aligning the two in one representation lets each constrain what the other leaves ambiguous, so the joint description is richer than either alone. This is \emph{multimodal} learning~\cite{baltruvsaitis2019multimodal}, of which alignment is one core idea; a \emph{modality} is one kind of measurement, an image, a spectrum, a light curve, or a table of scalar properties. The partners need not even be different kinds of data. Aligning a low-resolution and a high-resolution spectrograph that cover different wavelengths, but observe the same stars, is the same problem and works for the same reason, since both measure one physical object and a shared representation can hold the parameters they agree on while setting aside what is particular to each.

Multimodal learning usually needs paired data, observations known to be of the same object, so the model learns what the modalities actually share rather than coincidences of the training set. Even then the alignment is hard, because modalities differ in dimensionality, noise, resolution, and the information they carry. Two encoders can be trained to use the same representation space and still fail to put matching objects near one another. This mismatch is called the \emph{modality gap}~\cite{liang2022modalitygap}.

There are two broad ways to combine modalities. One keeps a heavy, separate encoder for each and learns only how to relate their outputs, either by pulling paired inputs (e.g., an image and a spectrum of the same object) toward nearby points in a shared space, as CLIP does~\cite{radford2021clip}, or by linking largely pretrained encoders through a small learned link that controls how much to draw from each~\cite{alayrac2022flamingo}. The other trains a single network jointly over all modalities, after turning each input into a shared format that the network can process~\cite{mizrahi2023fourm,chameleon2024}. The first approach keeps each encoder modular and independently reusable, and costs less to train. In contrast, the second shares one network across modalities, so it can generate one modality from another and fill in a missing one, at the price of training everything together. Either way, combining modalities can lose accuracy that each had on its own~\cite{huang2022modalitycompetition,wang2020hard,zhang2024vlmclassifier, zhu2024conflict}, so the join has to be designed rather than assumed.

Whether the effort pays off depends on what the other modalities adds. Two complementary modalities each carry something the other lacks, so aligning them sharpens the representation~\cite{yuan2021multimodal}; redundant ones may mostly beat down noise without opening a new physical direction, and aligning to a noisier or biased partner can even do harm. Useful alignment is therefore more than feeding several inputs to one network. The representation has to record which observations belong to the same object, what the modalities share, and what belongs only to one of them. Forcing every direction of the two representations to agree can erase the modality-specific part. Conversely, leaving them unlinked blocks any cross-modal transfer. The aim is to keep both the shared physical factors and the extra information each instrument provides.

\begin{figure}[t]
\centering
\includegraphics[width=\textwidth]{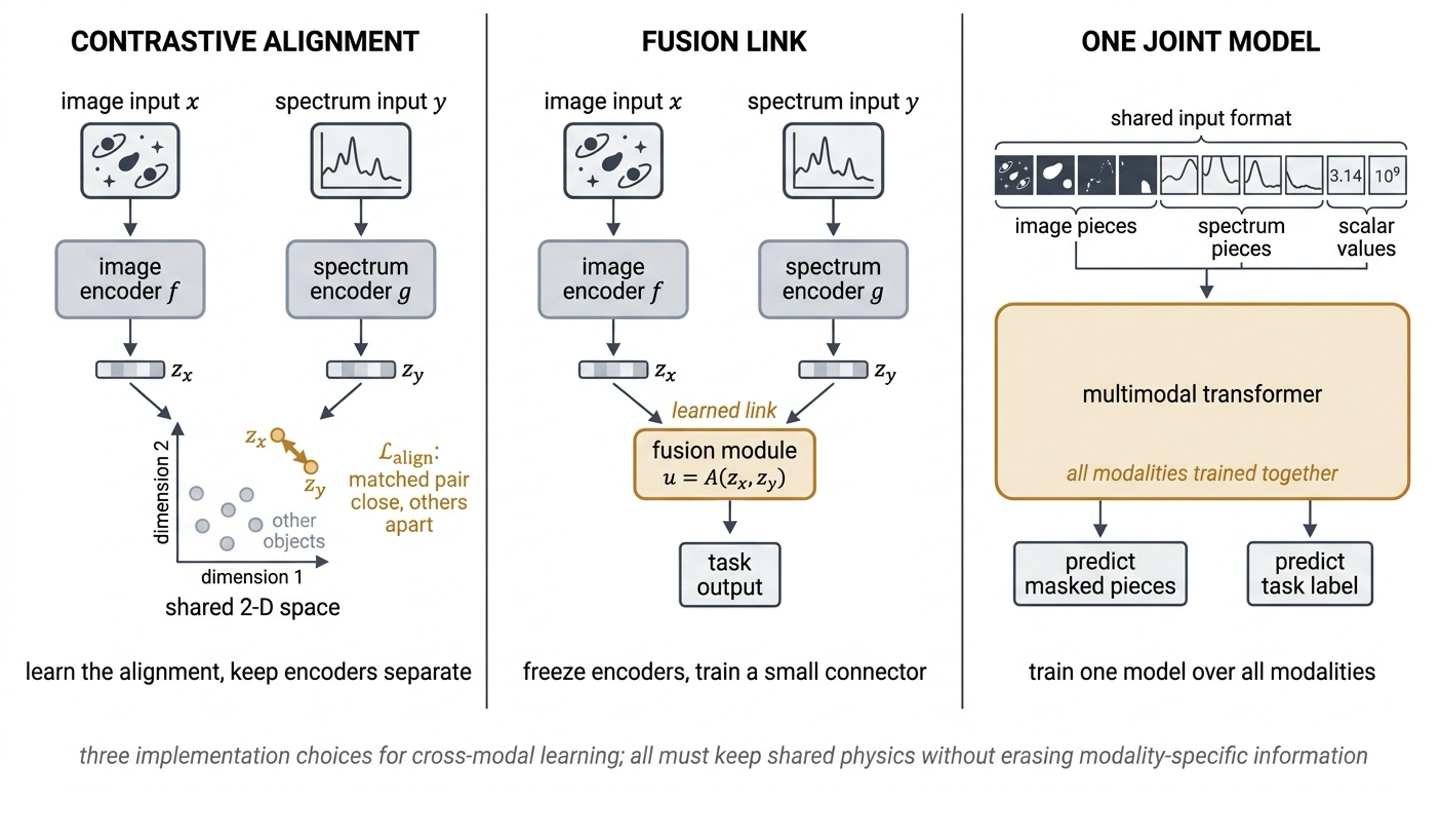}
\caption{Strategies for combining modalities. Separate encoders can be aligned by a contrastive loss, joined by a small learned connector, or replaced by a single model trained over a shared input format. In all cases, useful combination must keep the shared physical factors without erasing information carried only by one modality.}
\label{fig:cross_modal}
\end{figure}

Figure~\ref{fig:cross_modal} separates the implementation choices in this paragraph. Separate encoders can be aligned directly with a contrastive loss, connected through a small learned fusion module, or replaced by one joint model trained over all modalities. All three still face the same tension: the representation has to align what the modalities share without erasing what only one of them sees.

\subsection{Scaling and emergence}

\begin{figure}[t]
\centering
\includegraphics[width=\textwidth]{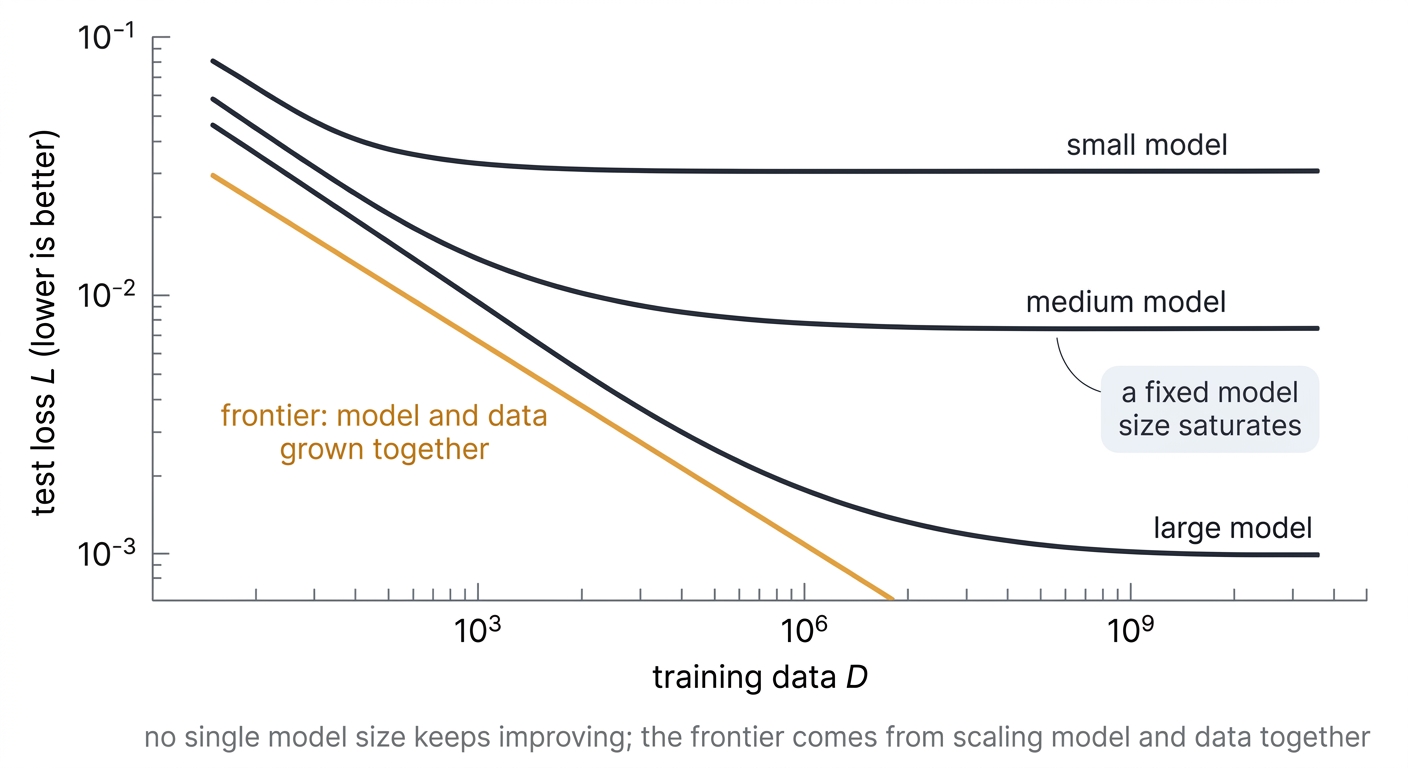}
\caption{Neural scaling. As the model and the training data grow, the test loss $\mathcal{L}$ falls as a power law, a steady and predictable improvement seen empirically across many settings rather than derived from first principles. Each charcoal curve is a model of fixed size trained on more data $D$, its loss falling and then flattening onto a plateau set by the model's capacity, so a small model saturates early and high while larger models saturate later and lower. The best loss at any data budget lies on the \emph{frontier} (amber), reached only by enlarging model and data together~\cite{kaplan2020scaling,hoffmann2022chinchilla}. Astronomy is usually data-limited, so this frontier is hard to climb, and a foundation model that leans on scale alone rather than on the structure of the data is liable to stall.}
\label{fig:scaling}
\end{figure}

Breadth across data types is one lever; sheer scale is another. Three quantities set how far a model can be pushed, the number of parameters $N$, the size of the training set $D$, and the compute $C$. As they grow, the test loss falls as a power law, roughly $\mathcal{L}\propto N^{-\alpha}$ in model size and likewise in data and compute, down to an irreducible floor $\mathcal{L}_{\infty}$ that more of either cannot beat; these regularities are known as \emph{neural scaling laws}~\cite{kaplan2020scaling} (Figure~\ref{fig:scaling}). The catch is that each axis saturates once another becomes the bottleneck, so a larger model on a fixed dataset soon stops improving. Model size and data therefore have to grow together, and a fixed compute budget is spent best by enlarging both in step rather than either one alone~\cite{hoffmann2022chinchilla}. Scaling became a deliberate strategy because this improvement is predictable and, past a certain size, can bring qualitatively new behaviour, the emergent abilities\footnote{The apparent suddenness of an emergent ability is in part an artifact of scoring~\cite{schaeffer2023mirage}, though not wholly so~\cite{du2024loss}.} that grokking shows in miniature.

Realizing these gains is also an engineering problem. The promised improvement follows only if training stays stable as the model grows, which takes systematic choices of optimizer, learning-rate schedule, and weight initialization, together with methods for carrying hyperparameters (configuration settings not learned from the data, such as learning rates or layer widths) tuned on small models up to large ones without repeating a costly search at every size. For stellar spectra these trade-offs have been mapped out directly, with scaling laws measured across model size, training-set size, and training length, together with a recipe for stable training and for transferring hyperparameters across model sizes through maximum-update parametrization (a method that scales learning rates and initialization so training dynamics remain stable as the network size grows)~\cite{rozanski2025scaling,Yang2020muP}. These methods are technical but not incidental, and whether an astronomical foundation model reaches the accuracy its size and data allow can depend on them as much as on the architecture.

Scale is a weaker lever in astronomy than in language. Real observations are many but nowhere near the scale of the largest models, and how many a survey gathers is bounded by the sky and the available telescopes. Simulations can enlarge the training set past that limit, and models are often trained on them, but realistic ones are themselves costly to run and bring the synthetic--observed gap with them, so data, observed or simulated, stays a real constraint. Where data rather than compute binds, the architecture and the self-supervised objective, and the assumptions they build in, matter more than scale alone, because an inductive bias that encodes real structure is information the model does not have to learn from data.

\subsection{Fine-tuning and adaptation}

Pretraining produces a representation, but most scientific use needs a specific \emph{label} read off it, a number to regress such as a redshift or a stellar parameter, or a class to assign such as a morphological type. The model is turned to that task by \emph{fine-tuning} it on the labelled examples available.\footnote{Language models add a further, reinforcement-learning stage on top of supervised fine-tuning, using human preferences between answers and related methods to make the model follow instructions and show its reasoning~\cite{ouyang2022instructgpt,rafailov2023dpo,wei2022cot,guo2025deepseekr1}. For a measurement model the supervised step is the one that matters.} Write the network as $f_{\theta}$, where $\theta$ collects all of its adjustable weights, arranged internally as a stack of weight matrices, and write its representation as $\mathbf{h}=f_{\theta}(\mathbf{x})$. The first choice is how many of those weights to move, from none to all (Figure~\ref{fig:adaptation}):

\begin{figure}[t]
\centering
\includegraphics[width=\textwidth]{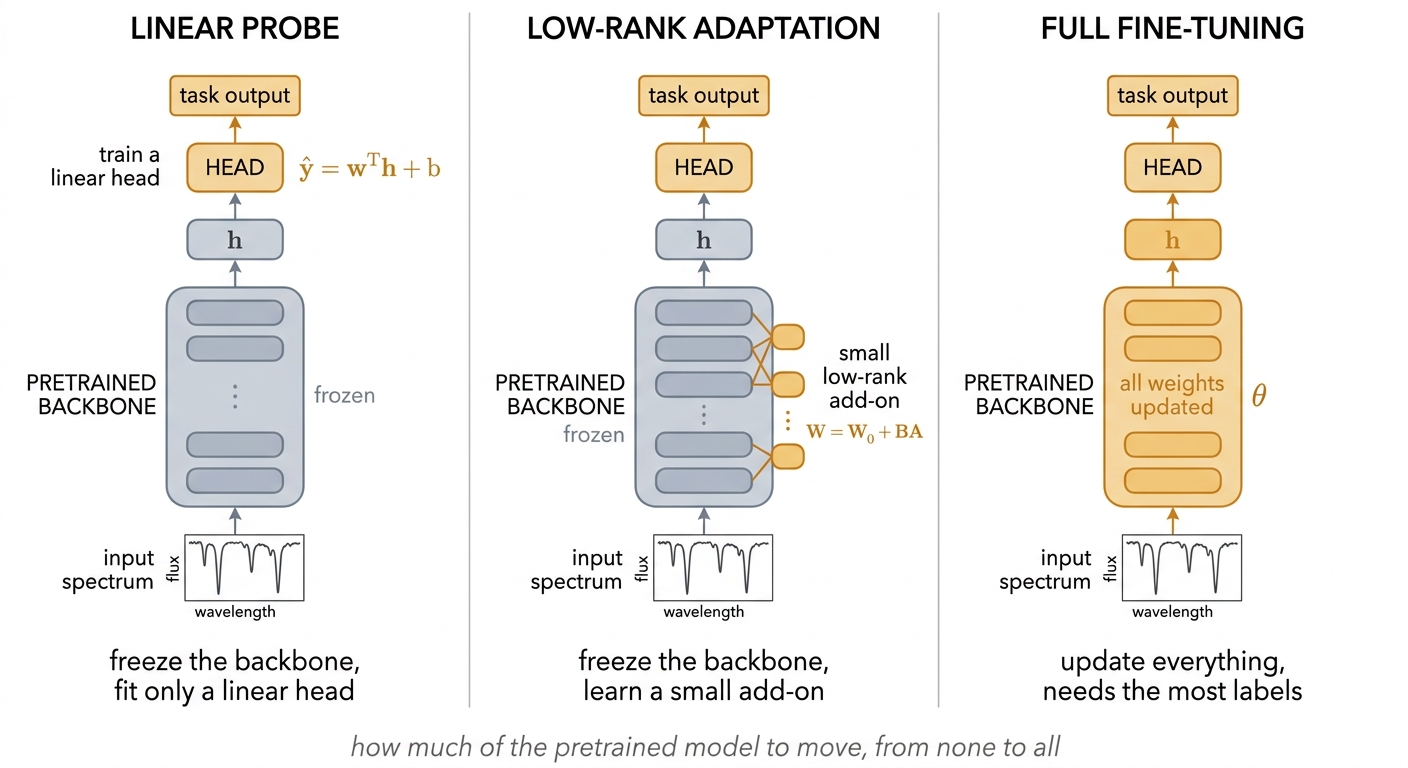}
\caption{Three ways to adapt a pretrained model to a task, ordered by how many weights move. A \emph{linear probe} freezes the pretrained network (the \emph{backbone}) and trains only a small output layer, in the simplest case linear, $\hat{y}=\mathbf{w}^{\top}\mathbf{h}+b$. \emph{Low-rank adaptation} keeps the backbone frozen and learns a small add-on, $\mathbf{W}=\mathbf{W}_0+\mathbf{B}\mathbf{A}$, with far fewer parameters. \emph{Full fine-tuning} updates all of the weights $\theta$, fitting the most but needing the most labels. Frozen parts are shown in grey, trained parts in amber.}
\label{fig:adaptation}
\end{figure}
\begin{itemize}
\item A \emph{linear probe} leaves $\theta$ frozen and trains only a small output layer on top of $\mathbf{h}$, in the simplest case a linear one, $\hat{y}=\mathbf{w}^{\top}\mathbf{h}+b$, with a weight vector $\mathbf{w}$ and offset $b$. It works when the label is already close to a linear function of $\mathbf{h}$, as when a star's temperature can be read straight off a frozen spectral embedding. A small nonlinear output layer is sometimes used instead when the aim is task performance rather than a clean readout of what the representation already encodes linearly.
\item \emph{Low-rank adaptation} moves a little more, but cheaply. It freezes a large pretrained weight matrix $\mathbf{W}_{0}$ and learns only a small add-on in its place, $\mathbf{W}=\mathbf{W}_{0}+\mathbf{B}\mathbf{A}$, where the add-on $\mathbf{B}\mathbf{A}$ is built from two much smaller matrices, so that its rank $r$ stays far below the matrix width $d$. That leaves about $d/2r$ times fewer numbers to fit than $\mathbf{W}_{0}$ holds, so a large model can be steered to a new survey with far fewer trainable parameters~\cite{hu2021lora}.
\item \emph{Full fine-tuning} updates all of $\theta$, every weight matrix at once, reshaping the representation around the label. It can fit the most but needs the most labels and can overwrite what pretraining built.
\end{itemize}

A second choice is when the task is trained relative to the self-supervised objective:
\begin{itemize}
\item the \emph{sequential} schedule first minimizes the self-supervised loss $\mathcal{L}_{\mathrm{SSL}}$ alone, then fits the task loss $\mathcal{L}_{\mathrm{task}}$ on the frozen or further-tuned result;
\item the \emph{joint} schedule minimizes the two at once, $\mathcal{L}=\mathcal{L}_{\mathrm{SSL}}+\lambda\,\mathcal{L}_{\mathrm{task}}$, with the weight $\lambda$ setting how hard the labels pull on the representation.
\end{itemize}
Joint training can help when labels are scarce, because the self-supervised loss keeps the representation tied to the structure of the unlabelled data while the labelled loss steers it toward the task~\cite{Walmsley2022jointschedule}. Its cost is that the two losses have to be balanced. If the task loss dominates, the representation may become too task-specific, while if the self-supervised loss dominates, the labels may not shape it enough~\cite{Kendall2017multi,Sener2018multi}. In the label-scarce setting of astronomy, where target datasets often differ from training datasets (a problem known as \emph{distribution shift}), the lighter output layers and the sequential schedule remain the cheaper choice and the cleaner test of whether the representation has captured transferable physics.

\section{Foundation-model attempts in astronomy}

The architectures, self-supervised pretraining, scaling, and fine-tuning set out so far were developed for language and, to some extent, vision~\cite{dosovitskiy2021vit, radford2021clip, caron2021dino, oquab2023dinov2, alayrac2022flamingo}. Astronomical data resemble neither. A spectrum, a light curve, or an image of a celestial object is produced by physical processes rather than by human communication, so its content is fixed by physics rather than by grammar or by the makeup of an everyday scene, and the labels that would interpret it are scarce. What carries over is not the subject matter but the goal, a representation that captures the physical factors behind an observation and stays useful when the instrument, the population, or the task changes. The attempts below borrow the machinery and ask whether it delivers.

\subsection{Bridging simulations and observations}

Before any of these models can be trusted, astronomy has to face a problem that language and vision do not. Trustworthy labels, the physical parameters behind an observation, are scarce and costly, while physics-based simulators can generate labelled synthetic spectra, images, or light curves in regimes where observed labels are sparse. The temptation is therefore to train on simulations and let the labels come cheaply. The catch is that a model trained on synthetic data alone learns the simulator as much as the physics, its line list or sub-grid recipe, the instrument's resolution and noise, the calibration, and the particular population the parameters were drawn from. On real data all of these differ, and this \emph{domain gap} (the discrepancy between simulated and real data) shifts both the inputs and their relation to the labels, so it cannot be papered over by adding random noise. Several strategies are deployed against it:
\begin{itemize}
\item \emph{forward modeling} passes the simulation through an instrument and noise model before training, so the network sees synthetic data dressed to look observed;
\item \emph{data augmentation} varies the known observational effects so the representation does not lock onto one realization of them;
\item \emph{domain adaptation} learns a mapping between the synthetic and observed distributions, as \textsc{Cycle-StarNet} does by translating theoretical spectra into the observed domain without paired counterparts~\cite{OBriain2021cyclestarnet};
\item \emph{synthetic pretraining} trains on abundant simulations and then calibrates on a smaller observed set, as \textsc{SpectraFM} does across wavelength ranges ~\cite{koblischke2024spectrafm}, the same logic as multifidelity emulation, where many cheap, lower-fidelity runs are corrected by a few expensive ones~\cite{ho2022multifidelity,Saoulis2025transfercosmos}.
\end{itemize}
None of these guarantees that a model trained on simulations will behave on the sky, so each has to be tested on real data held out by instrument, population, or signal-to-noise, not on a random split of the training mixture.

\subsection{Models by modality}

Each modality borrows the recipe whose built-in assumptions fit it best, since the structure of the measurement decides which self-supervised objective is natural. A spectrum or a light curve is a sequence, like the tokens in a sentence, making masked and next-step objectives a natural fit. A survey image plays the part of a natural image, suited to the augmentation-based and patch-based objectives of vision. Finally, two instruments observing one object play the part of an image and its caption, forming a matched pair for contrastive alignment. Across the modalities the same pattern recurs, a recipe carried over from language or vision, a science target set by scarce labels, a handful of pretrained models, and growing evidence of neural scaling laws (Table~\ref{tab:astrofm}).

In each case the frozen representation is then put to work on a scientific task. A small output layer regresses a physical parameter or assigns a class. The geometry of the embedding supports nearest-neighbour search and anomaly detection, although, since instrumental faults, processing failures, and new physics all surface as outliers, an anomaly flag begins a triage rather than a discovery. The representation can also serve as a compact \emph{summary statistic} for \emph{simulation-based inference}, which replaces a likelihood too costly to evaluate with direct comparisons between data and simulations and can fold in generative models as learned priors or posterior samplers~\cite{cranmer2020sbi,Ho2024ILI,ono2024debiasing}. Such inference is trustworthy only where its posterior coverage survives a change of nuisance parameters, instrument, or simulator, not merely where the generated data look realistic.

\begin{table}[H]
\caption{Foundation-model attempts in astronomy by modality, showing the data, the architecture, the self-supervised objective carried over from language or vision, representative models, and the science targets.}
\label{tab:astrofm}
\renewcommand{\arraystretch}{1.2}
\scriptsize
\begin{tabular}{@{}L{1.55cm}L{1.75cm}L{1.35cm}L{1.8cm}L{2.05cm}L{1.85cm}@{}}
\hline\noalign{\smallskip}
\textbf{Modality} & \textbf{Data} & \textbf{Architecture} & \textbf{Objective} & \textbf{Models} & \textbf{Targets} \\
\noalign{\smallskip}\svhline\noalign{\smallskip}
Time domain & Light curves (\textit{Kepler}) & Transformer & Masked, autoregressive & ASTROMER~\cite{donoso2023astromer}, FALCO~\cite{zuo2025falco} & Variability, transients, anomaly \\
Spectroscopy & Stellar spectra (\textit{Gaia}~XP, DESI) & Transformer & Masked, autoregressive, contrastive & Leung \& Bovy~\cite{leung2024stellarfm}, SpecCLIP~\cite{zhao2025specclip} & Parameters, abundances, redshift \\
Imaging & Survey images & CNN, transformer & Contrastive, autoregressive & Hayat et al.~\cite{hayat2021ssl}, AstroPT~\cite{smith2024astropt} & Morphology, photo-$z$, search \\
Multimodal & Image, spectrum, scalars & Transformer & Contrastive, tokenized masking & AstroCLIP~\cite{parker2024astroclip}, AION-1~\cite{parker2025aion} & Cross-modal estimation \\
Other modalities/messengers & Radio, grav.\ waves & Varies & Self-supervised & SKATR~\cite{ore2024skatr}, GWAK~\cite{raikman2023gwak} & Source analysis, inference, anomaly \\
Physical simulations & PDE, hydrodynamic fields & Transformer & Cross-system pretraining & MPP~\cite{mccabe2023mpp}, Walrus~\cite{mccabe2025walrus} & Surrogate, transfer across systems \\
\noalign{\smallskip}\hline
\end{tabular}
\end{table}

\paragraph{Time domain.} Time-domain astronomy studies how sources change, and its basic measurement is the \emph{light curve}, the brightness of a source recorded over time, which modern surveys gather for millions of objects at once. Its tasks are the ones where labels are scarce but the data overwhelming, classifying periodic variable stars, catching transients such as supernovae, picking out active galactic nuclei and quasars by their irregular flickering, detecting flares, and flagging anomalies that might be something new. A light curve is a sequence, so the masked and next-step objectives apply, with one complication, that astronomical sampling is irregular and each point carries its own uncertainty; the observation times and errors are therefore fed to the network alongside the brightnesses, so an uneven cadence becomes information rather than missing data.

\textsc{ASTROMER} pretrains a transformer on large sets of unlabelled light curves in this way and is then fine-tuned to classify variable stars from only a few labels~\cite{donoso2023astromer}, while \textsc{FALCO} applies the same recipe to \textit{Kepler} space photometry, handling variability classification, surface-gravity estimation, and flare detection from a single pretrained model~\cite{zuo2025falco}. The predictable improvement with scale that drives language models appears here too, with a light-curve model obeying the same neural scaling law as the training set, the model, and the compute grow~\cite{Pan2024scalinglaw}. And when a source is also observed another way, its light curve can be aligned with a spectrum or with catalogue metadata so that each sharpens the other~\cite{zhang2024maven,rizhko2024astrom3,Kamai2025dual}, towards multimodal models.

\paragraph{Spectroscopy.} Spectroscopy spreads a source's light by wavelength, and for a star the resulting spectrum is set by a handful of physical quantities, its temperature, surface gravity, and chemical composition, acting through known atomic physics. That makes spectra a particularly clean case for a transferable representation, because the factors of variation really are few and physical. It also makes the natural targets a set of regressions, including effective temperature, surface gravity, overall metallicity, the abundances of individual elements, radial velocity, and, for galaxy spectra, redshift.

Masked and generative pretraining carry over as for any sequence, and a single transformer trained on \textit{Gaia}~XP spectra, the very low-resolution spectra of hundreds of millions of stars from the \textit{Gaia} satellite, can derive these parameters, generate a spectrum from given labels, and fill in masked wavelength regions, all from one model~\cite{leung2024stellarfm}. The harder and more revealing goal is to relate different instruments, since a star observed by two spectrographs should map to the same parameters even though the raw spectra differ in resolution and wavelength coverage. Joint-embedding methods pursue this by aligning spectra of the same stars from different surveys into one shared space~\cite{Tobias2024XPRVS,zhao2025specclip}, and \textsc{SpecCLIP} has been carried to the DESI spectroscopic survey from only a handful of labels through low-rank fine-tuning, though its accuracy degrades in the sparse tails of the label distribution~\cite{zhao2025lora}. A complementary line unifies at the input rather than the embedding: a single encoder ingests heterogeneous spectra on their native wavelength grids, without resampling or cross-matched pairs~\cite{shen2025tokenizer,islam2026omnispectra}. Spectral models obey the same neural scaling law~\cite{rozanski2025scaling}.

\paragraph{Optical imaging.} Optical imaging surveys record the sky as pictures, and the self-supervised recipes of computer vision transfer most directly here. The standard approach makes two augmented views of one galaxy, by rotating, cropping, or adding noise, and trains the network so that the two views land together, either by also pushing different galaxies apart, the contrastive recipe of SimCLR~\cite{chen2020simclr}, or by a self-distillation scheme (which avoids the need for negative examples by training the network to match its own predictions), as in DINO~\cite{caron2021dino}. Alternatives mask image patches and rebuild them~\cite{he2022mae}, or predict them in sequence~\cite{el2024aim}. Either way the network learns without labels, provided the augmentations leave the target untouched, which is itself a modelling choice.

The science targets are morphological classification, photometric redshift, and large-scale search, and a frozen encoder reaches supervised accuracy on morphology and redshift from a small labelled set~\cite{hayat2021ssl,parker2024astroclip}. Because the embedding places similar galaxies near one another, the same features also drive similarity searches and hunts for rare objects such as gravitational lenses across tens of millions of images~\cite{stein2022lenses}. \textsc{Fama} scales a masked autoencoder into a foundation model that transfers across surveys and downstream tasks~\cite{lv2026fama}. \textsc{AstroPT} takes the autoregressive route, predicting image patches in sequence across millions of galaxies, and obeys the same neural scaling law~\cite{smith2024astropt}.

\paragraph{Multimodal.} When several instruments observe one object, a single backbone can be trained to span them, so that a property can be read from whichever measurement is in hand. Unlike the preceding examples that combine light curve, (multi-instrument) spectra, and metadata~\cite{zhang2024maven,rizhko2024astrom3,Kamai2025dual,Tobias2024XPRVS,zhao2025specclip}, \textsc{AstroCLIP} aligns galaxy images with spectra into one embedding space, in the manner of CLIP, and then supports retrieval and prediction from either input~\cite{parker2024astroclip}. \textsc{AION-1} goes further, converting images, spectra, and scalar measurements into a shared input format, in the manner of the multimodal model \textsc{4M}~\cite{mizrahi2023fourm}, and training one transformer to predict masked pieces of any modality. A small output layer on the frozen backbone can then handle many tasks at once~\cite{mizrahi2023fourm,parker2025aion}, with related designs differing in how the continuous measurements are encoded~\cite{shen2025universal}.

\paragraph{Other modalities and messengers.} Beyond optical imaging and spectroscopy the coverage thins quickly, though the recipe has been carried to other modalities and messengers. In radio astronomy, a self-supervised backbone learns from survey images of radio galaxies and supports their morphological classification~\cite{slijepcevic2024radioFM}. In cosmology, the \textsc{SKATR} summary transformer compresses simulated 21\,cm intensity maps into features for downstream parameter inference~\cite{ore2024skatr}. For gravitational waves, \textsc{GWAK} detects anomalies with recurrent autoencoders, networks that read the strain in time order, flagging candidate signals without templates~\cite{raikman2023gwak}. Each shows the recipe porting to a new modality or messenger, but none has yet been reused \emph{across} messengers, the cross-domain transfer a true backbone would demonstrate.

\paragraph{Physical simulations.} A separate line trains not on observations of the sky but on the output of physics simulations, such as hydrodynamic and other partial-differential-equation solvers, distinct from the synthetic spectra and images used as labels elsewhere. Multiple-physics pretraining trains one transformer on the solutions of many different equations at once, so that it transfers to a physical system it was not trained on~\cite{mccabe2023mpp}, and \textsc{Walrus} extends the idea to a broad range of continuum-dynamics problems~\cite{mccabe2025walrus}. This is the closest the field has come to a model that has learned transferable physics, though so far in the controlled setting where the governing equations are known and the data are clean.

\subsection{Limitations}

Set against the progress above are several reasons the foundation-model label still outruns the evidence in astronomy. The root difficulty is that a high score on held-in data need not mean transfer. A feature that predicts metallicity, for instance, might encode the absorption lines of those elements, a temperature correlation that merely happens to hold in the training set, or survey-specific calibration; all three lift the in-distribution score (the score on data drawn from the training population), but only the first survives a change of population or instrument. A real test therefore holds out a whole instrument, population, or signal-to-noise range, rather than a random subset, and asks whether both the accuracy and the calibrated uncertainty survive. Several recurring obstacles stand in the way.

\paragraph{Few clear wins over supervised baselines.} The promise of a foundation model is to do more with fewer labels, yet where labels are plentiful a pretrained model often fails to beat a supervised one trained directly on the task. Reported gains also often lack a matched baseline, meaning the comparison should change only the pretraining, while keeping the labelled set, the small task-specific output layer, and the evaluation fixed~\cite{ting2026deeplearning}. The comparison is muddied further because many reference labels are themselves outputs of a physical-model fit, so a network that reproduces them is rewarded for emulating the pipeline, including its systematics, rather than for recovering the underlying physical quantity. The honest case for these models is the label-poor regime; where labels are abundant, direct supervision remains the baseline to beat~\cite{Lastufka2025supervisedcompare}.

\paragraph{The synthetic gap is rarely closed.} Because labels come most readily from simulations, the natural recipe is to pretrain on synthetic data and apply the model to the sky. But the bridging strategies of the opening section are seldom shown to carry across the gap on real data held out by instrument or population, as opposed to a random split, so direct evidence that a representation transfers from simulation to observation stays thin.

\paragraph{Inductive biases may not fit the data.} A transformer's built-in assumptions are not automatically the right ones for a spectrum or an image. On some spectral tasks a plain fully connected network has been reported to transfer across resolutions well~\cite{Zhao2026generalize}, and spectral emulators remain data-hungry rather than data-efficient, their error still falling steeply with more data~\cite{rozanski2025scaling}. Either the assumed bias is mismatched to the data, or it is not yet robust, or the training sets are simply not large enough to pay for the flexibility a transformer brings; the cases so far do not separate these explanations.

\paragraph{Tokenization has no natural answer.} A transformer reads a sequence of units, or \emph{tokens}. In language those units come naturally as words or word pieces, but astronomy does not supply them ready-made. A spectrum or an image is continuous, so current models either sidestep the question with fully connected networks, cut the input into patches, or learn a codebook (a discrete dictionary of learned vector patterns) with no physical meaning, any of which can blunt the architecture's advantage and may be part of why the inductive bias above fails to pay off. A tokenization tied to the physics, splitting a spectrum into its line profiles rather than fixed bins, might transfer better but is so far essentially unexplored.

\paragraph{Reconstruction is a weak pretext (surrogate) task.} Almost all of the pretraining above rests on reconstruction, predicting masked or missing values, which rewards getting every pixel right. The physics, though, is usually carried by a few localized features, a handful of absorption lines against a bright continuum, so a reconstruction loss is dominated by the easy, high-variance bulk and need not track the faint diagnostic signal~\cite{ting2026deeplearning}. The trade-off is the one principal-component analysis makes plain, where the directions of greatest variance are not always the ones that carry the label of interest. A pretext task aligned with the physical structure of the data, rather than with raw reconstruction, is still missing, and is among the clearer openings for progress.

\section{Open questions and future directions}

The distinction that runs through this chapter is between the goal and the means. The architectures, self-supervised objectives, scaling, cross-modal alignment, and adaptation gathered here were worked out for language and carried into astronomy largely intact, yet assembling them is not the same as building a foundation model. A transformer trained and evaluated on one task is a task-specific model, however large; what earns the name is evidence that its representation survives a change of task, instrument, or population, through zero-shot or few-shot use or cross-instrument transfer. That evidence, not the architecture, is what the term should denote, and across astronomy it remains scarce. These open questions are about the distance between the toolkit and that goal.

The gap is not peculiar to astronomy, because the language success has not carried cleanly to other domains. In vision, once an image is cut into patches the same self-supervised objectives apply, through masked reconstruction~\cite{he2022mae}, contrastive alignment between two augmented views of one image~\cite{chen2020simclr} and MoCo~\cite{he2019Moco}, self-distillation~\cite{caron2021dino,oquab2023dinov2,grill2020byol}, auto-regressive generation~\cite{el2024aim}, or alignment with text~\cite{radford2021clip}, yet no single visual model yet spans the full task range that one language model handles, vision still leaning on separate models for segmentation (identifying the boundaries of individual objects), generation, and the rest~\cite{kirillov2023sam,rombach2022ldm,awais2024visionfm}. A model that generates a convincing image has learned to produce samples from the data distribution, which does not thereby yield a representation whose coordinates support measurement, so a visually plausible galaxy need not preserve flux, morphology, or uncertainty. Vision-language models join image and text but have not yet produced the robust general understanding once hoped for, and self-supervised video models remain less mature~\cite{Schiappa2022videosurvey}. Fluent generation has outpaced reliable understanding.

For astronomy the question is sharper, because it asks for physical rather than linguistic intelligence. Many astronomical models called foundation models are, at present, transformers pretrained and evaluated within a narrow regime, and demonstrations of the transfer the term denotes remain uncommon, which is less a shortcoming of the work than an expectation that was never warranted. Physical inference is harder than the settings where the recipe succeeded, since the inputs are governed by physics, the labels come from imperfect models, and the standard is not fluent generation but calibrated, transferable measurement. Whether the way forward is more of the present recipe at greater scale~\cite{Huh2024platonic,UniverseTBD2025platonic,smith2024astropt} or a different account of what a good representation is~\cite{lecun2022path} remains unsettled. One position holds that scaling will keep improving and eventually transfer. The other holds that compression with the methods collected here is not yet a theory of representation, and that going beyond language may take a new idea.

Whichever way that goes, the evidence the term demands has to be built into the evaluation rather than read off a leaderboard. A random split between training and test data preserves a survey's selection, its depth, cadence, targeting, and quality cuts, so a representation can score well by leaning on correlations that are merely instrumental. A real test of transfer instead holds out a whole sky region, time span, instrument, population, or range of signal-to-noise, with the sparsely sampled tails treated separately, since a large distance in embedding space can mean either a rare object or an ordinary one seen under unusual conditions. Three claims are easily run together and worth keeping apart. These are accurate prediction, transfer that survives such a change, and evidence that the representation tracks physical structure. Only the last two are what the foundation-model label is meant to add.

The last is the hardest to establish, because nothing in a prediction loss forces one coordinate of the representation to be temperature and another metallicity. The model could mix those coordinates together, and a later readout could unmix them while giving the same prediction, so the individual axes of the representation would not have a unique physical meaning. Unsupervised objectives face a fundamental mathematical limit. Without built-in physical assumptions, there are infinitely many ways to combine coordinates that reproduce the data equally well, meaning the network can easily mix physical factors together in a way that is mathematically correct but physically uninterpretable~\cite{locatello2019challenging}. Physical meaning therefore has to be earned with labelled standards, simulated interventions, known symmetries, or controlled changes of one factor at a time. That it is hard to establish does not necessarily make it costly to demand: interpretability need not come at the cost of accuracy, despite the two often being posed as a strict tradeoff~\cite{rudin2019stop}. Underlying all of this, the provenance of data, labels, simulations, and model weights has to be documented, since releasing the weights alone does not reproduce the data selection and optimization history that produced a model.

\subsection{Research directions}

\begin{figure}[t]
\centering
\includegraphics[width=\textwidth]{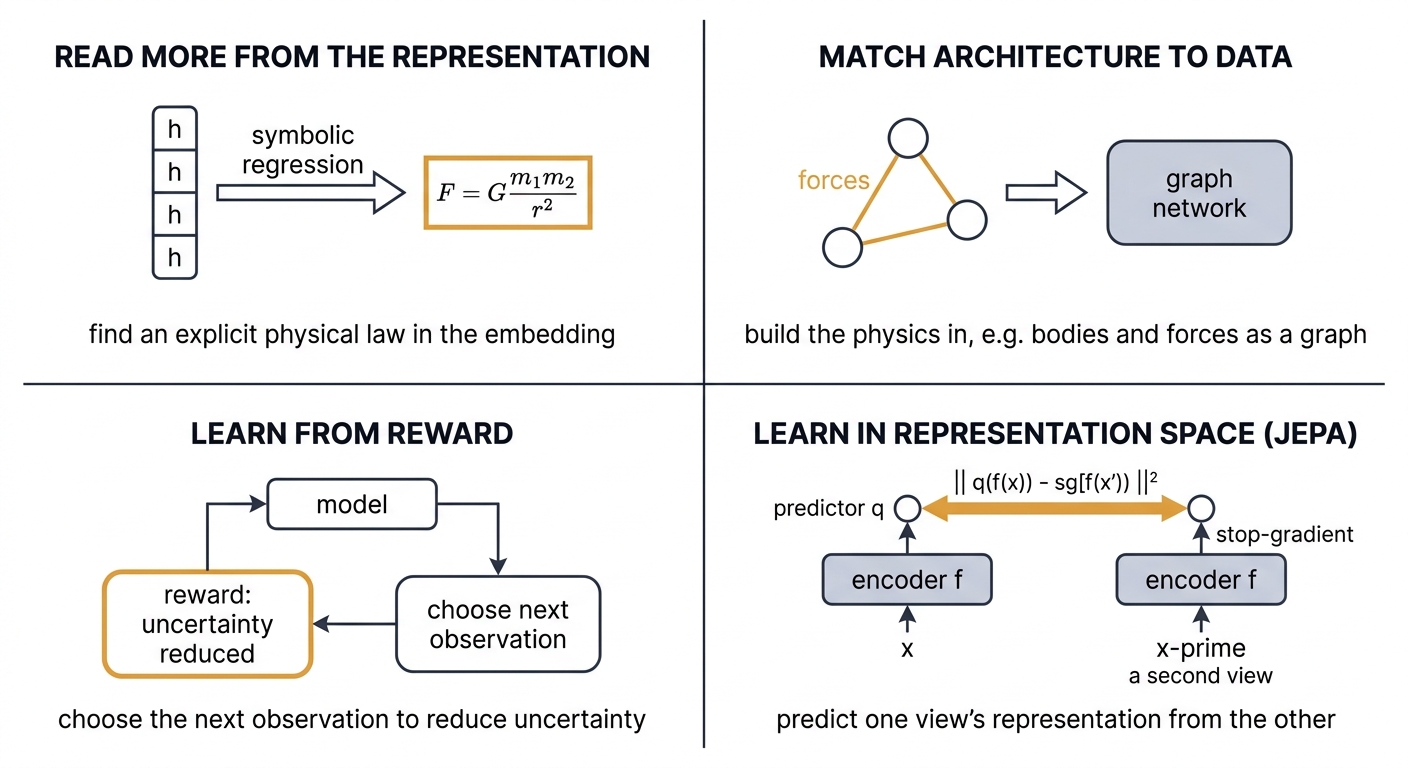}
\caption{Four research directions for astronomical foundation models, sketched as mechanisms. \emph{Reading more from the representation} extracts an explicit physical law from an embedding through symbolic regression. \emph{Matching the architecture to the data} builds the physics in, as a graph of bodies and forces. \emph{Learning from reward} closes a loop that chooses the next observation to reduce uncertainty. \emph{Learning in representation space} (JEPA) predicts one view's representation from another, comparing a predicted representation $q(f(\mathbf{x}))$ with a target representation held fixed during that update, $\mathrm{sg}[f(\mathbf{x}')]$. Each is worth only as much as the evidence that it transfers.}
\label{fig:directions}
\end{figure}

The limitations are also a map of where progress might come from, and a few directions take direct aim at them, running from the most grounded, making better use of the representations already in hand, to the most speculative, changing how those representations are learned. In each what matters is less the idea than the evidence that would show it had worked, since a method never held out against a change of instrument or population proves little about transfer (Figure~\ref{fig:directions}).

\paragraph{Reading more from the representations we have.} The most grounded direction supposes that the embeddings already learned carry more than has been read out of them, and sets out to understand them rather than to retrain. Symbolic regression searches for an explicit relation among embedding coordinates and physical quantities, and has recovered force laws, an analytic dark-matter relation, and an expression for the cosmic matter density~\cite{cranmer2020symbolic,lemos2023orbital,shao2023universal}, with tools such as AI Feynman, PySR, ESR, and PhySO~\cite{udrescu2020aifeynman,cranmer2023pysr,Bartlett2022ESR,PhySO_RL_DA,PhySO_ClassSR}, though any expression it returns must be checked on held-out regimes and against dimensional and limiting-case constraints. Read this way, a black-box encoder can be turned into something closer to a physical relation, often with no new training at all, only a sharper interrogation of what is already there.

\paragraph{Architectures matched to the data.} Architectural assumptions earn their keep in several ways in astronomy, when data or labels are scarce, a symmetry or conservation law fixed in the model is information it need not learn~\cite{ting2026deeplearning}; when they are abundant, the choice of structure still shapes what the model internalizes, not only what it predicts; and either way the right structure is the one matched to the data's form. Physics can be built into the architecture or loss through conservation laws, symmetries, partial differential equations, or differentiable simulators (simulators whose code allows automatic calculation of derivatives, enabling gradient-based optimization) that can be embedded in the training itself~\cite{greydanus2019hamiltonian,cohen2016group,karniadakis2021physics,raissi2019pinn,Li2025pinn,NEURIPS2018_842424a1}, which aids extrapolation when the assumed physics holds but can suppress the very effect one hoped to find if imposed too rigidly; the orbital case is telling, since a generic sequence model can predict trajectories without recovering Newtonian mechanics while a graph model that represents bodies and their pairwise forces makes the law recoverable~\cite{vafa2025worldmodel,lemos2023orbital}, a reminder that predictive equivalence is not mechanistic equivalence. The backbone itself is not settled, and state-space models, which process very long sequences at lower cost than attention, may suit high-resolution spectra and long light curves better than the transformer~\cite{gu2023mamba,gu2022s4}.

\paragraph{Learning from reward.} Supervised fine-tuning is not the only way to adapt a pretrained model. Large language models gained much of their later capability from reinforcement learning, in which a model is optimized not against fixed labels but against a reward, a stated measure of how good an output is, so that it learns from its own trials rather than from a supplied answer~\cite{ouyang2022instructgpt}, and pushing this further is the current frontier for language models. Its place in astronomy is less obvious, since most astronomical problems are measurement and inference, which have a right answer rather than a reward to be maximized. Where reward-based learning fits naturally is the engineering-optimization tasks, above all closed-loop experimental design, in which a model chooses the next target or exposure so as to reduce a stated uncertainty, with scarce telescope time as the resource to spend well~\cite{gilda2020imagequality}. How much further it will help a field built on measurement rather than action is, for now, unclear.

\paragraph{Learning in representation space.} The most speculative direction changes the pretraining objective itself. Reconstruction rewards predicting every value rather than capturing the few features that carry the physics, which is a reason to move the prediction off the raw data. Joint-embedding predictive architectures (JEPA) do exactly that by predicting the representation of a hidden or future part of the input, rather than reconstructing its pixels. This keeps the joint-embedding idea, but shifts the question from matching two augmented views to predicting what the unseen part should look like in representation space. Where an autoencoder minimizes $\lVert \mathbf{x} - g(f(\mathbf{x}))\rVert^{2}$ and a contrastive objective leans on negative examples, these methods minimize the distance between a predicted and a target representation, for instance $\lVert q(f(\mathbf{x})) - \mathrm{sg}[f(\mathbf{x}')] \rVert^{2}$, where $\mathbf{x}'$ is a second view, $q$ is a small predictor, and $\mathrm{sg}[\cdot]$ represents the stop-gradient operator (which treats the target representation as a constant during optimization to prevent collapse) that the predictor is trained to match~\cite{lecun2022path,assran2023ijepa,assran2025vjepa2,Balestriero2025lejepa}.

This is more than a better pretext (surrogate) task. A model that predicts how a scene changes in representation space is a step toward what is now called \emph{physical AI}, a model that anticipates how a scene responds to being acted on rather than only describing it, the ambition behind the vision-language-action models of robotics that map what is seen and instructed onto what to do next~\cite{Brohan2023RT2, Kim2024OpenVLA}. Astronomy has little use for the action half, but the representational objective, which keeps the structure that matters and lets go of the unpredictable detail, is what a transferable encoder needs.

\section{Concluding remarks}

Underneath the methods, what foundation models reach for is a working notion of machine intelligence, and its broadest form is what is loosely called a \emph{world model}, a model that has captured enough of how the world works to carry beyond the data it was trained on. The phrase holds many definitions and as many approaches, from predictive embeddings to generative simulators to agents that plan, and no single account has settled. For astronomy the concrete core of the idea is a representation that carries across instruments, populations, and tasks. Whatever shape a world model finally takes, one thing is clear, that astronomy will not reach it by scale alone, since it is a data-starved field without the means that drove the recipe in language, so the borrowed machinery applied unchanged is unlikely to be enough and progress will turn on innovation in how representations are learned.

The deeper ambition is larger than convenience. Astronomy has long read its data through heuristics built by hand, colours, line indices, fitted templates, each a compression chosen in advance. A representation learned from the data could in principle recover the compact description that the physics implies but no one has written down, and so capture structure that those heuristics miss. Present models fall well short of this, but it is the right aim, because it marks the difference between automating the measurements we already make and uncovering the variables we should have been measuring.

The exchange runs both ways. The settings where machine intelligence has advanced most are those with abundant data and forgiving standards, where fluent generation can pass for understanding. Astronomy offers the opposite, data whose every feature is set by physics, processes whose governing equations are partly known, and several instruments that view the same object, so that a claimed representation can be tested against physical law and across modalities rather than against human judgement. That makes the field not only a consumer of these methods but a proving ground for them, a place where science can serve AI as much as AI serves science.

The goal is reachable. The language models are an existence proof that transferable structure can be learned rather than merely hoped for, and doing the same in the physical world, where transfer is still rare, is the work ahead. It will be earned by holding the goal apart from the toolkit and demanding the evidence that separates a representation that has understood something from one that has only fit the data.

\begin{acknowledgement}

XZ is supported by a grant from the Schmidt Sciences. YST is supported by NSF under Grant AST-2406729 and a Humboldt Research Award from the Alexander von Humboldt Foundation. Copy-editing and literature research for this chapter were aided by the Claude-Opus-4.8 and GPT-5.5 large language models. The structure, ideas, and validation of this work are entirely due to the authors.
\end{acknowledgement}

\bibliographystyle{plain} 
\bibliography{references}

@article{olshausen1996emergence,
  title={Emergence of simple-cell receptive field properties by learning a sparse code for natural images},
  author={Olshausen, Bruno A and Field, David J},
  journal={Nature},
  volume={381},
  number={6583},
  pages={607--609},
  year={1996},
  publisher={Nature Publishing Group UK London}
}

@inproceedings{NEURIPS2018_842424a1,
 author = {de Avila Belbute-Peres, Filipe and Smith, Kevin and Allen, Kelsey and Tenenbaum, Josh and Kolter, J. Zico},
 booktitle = {Advances in Neural Information Processing Systems},
 editor = {S. Bengio and H. Wallach and H. Larochelle and K. Grauman and N. Cesa-Bianchi and R. Garnett},
 pages = {},
 publisher = {Curran Associates, Inc.},
 title = {End-to-End Differentiable Physics for Learning and Control},
 url = {https://proceedings.neurips.cc/paper_files/paper/2018/file/842424a1d0595b76ec4fa03c46e8d755-Paper.pdf},
 volume = {31},
 year = {2018}
}

@ARTICLE{Yang2020muP,
       author = {{Yang}, Greg and {Hu}, Edward J.},
        title = "{Feature Learning in Infinite-Width Neural Networks}",
      journal = {arXiv e-prints},
         year = 2020,
        month = nov,
          eid = {arXiv:2011.14522},
        pages = {arXiv:2011.14522},
          doi = {10.48550/arXiv.2011.14522},
archivePrefix = {arXiv},
       eprint = {2011.14522},
 primaryClass = {cs.LG},
       adsurl = {https://ui.adsabs.harvard.edu/abs/2020arXiv201114522Y}
}

@ARTICLE{Brohan2023RT2,
       author = {{Brohan}, Anthony and {Brown}, Noah and {Carbajal}, Justice and {Chebotar}, Yevgen and {Chen}, Xi and {Choromanski}, Krzysztof and {Ding}, Tianli and {Driess}, Danny and {Dubey}, Avinava and {Finn}, Chelsea and {Florence}, Pete and {Fu}, Chuyuan and {Gonzalez Arenas}, Montse and {Gopalakrishnan}, Keerthana and {Han}, Kehang and {Hausman}, Karol and {Herzog}, Alexander and {Hsu}, Jasmine and {Ichter}, Brian and {Irpan}, Alex and {Joshi}, Nikhil and {Julian}, Ryan and {Kalashnikov}, Dmitry and {Kuang}, Yuheng and {Leal}, Isabel and {Lee}, Lisa and {Lee}, Tsang-Wei Edward and {Levine}, Sergey and {Lu}, Yao and {Michalewski}, Henryk and {Mordatch}, Igor and {Pertsch}, Karl and {Rao}, Kanishka and {Reymann}, Krista and {Ryoo}, Michael and {Salazar}, Grecia and {Sanketi}, Pannag and {Sermanet}, Pierre and {Singh}, Jaspiar and {Singh}, Anikait and {Soricut}, Radu and {Tran}, Huong and {Vanhoucke}, Vincent and {Vuong}, Quan and {Wahid}, Ayzaan and {Welker}, Stefan and {Wohlhart}, Paul and {Wu}, Jialin and {Xia}, Fei and {Xiao}, Ted and {Xu}, Peng and {Xu}, Sichun and {Yu}, Tianhe and {Zitkovich}, Brianna},
        title = "{RT-2: Vision-Language-Action Models Transfer Web Knowledge to Robotic Control}",
      journal = {arXiv e-prints},
         year = 2023,
        month = jul,
          eid = {arXiv:2307.15818},
        pages = {arXiv:2307.15818},
          doi = {10.48550/arXiv.2307.15818},
archivePrefix = {arXiv},
       eprint = {2307.15818},
 primaryClass = {cs.RO},
       adsurl = {https://ui.adsabs.harvard.edu/abs/2023arXiv230715818B}
}

@ARTICLE{Kim2024OpenVLA,
       author = {{Kim}, Moo Jin and {Pertsch}, Karl and {Karamcheti}, Siddharth and {Xiao}, Ted and {Balakrishna}, Ashwin and {Nair}, Suraj and {Rafailov}, Rafael and {Foster}, Ethan and {Lam}, Grace and {Sanketi}, Pannag and {Vuong}, Quan and {Kollar}, Thomas and {Burchfiel}, Benjamin and {Tedrake}, Russ and {Sadigh}, Dorsa and {Levine}, Sergey and {Liang}, Percy and {Finn}, Chelsea},
        title = "{OpenVLA: An Open-Source Vision-Language-Action Model}",
      journal = {arXiv e-prints},
         year = 2024,
        month = jun,
          eid = {arXiv:2406.09246},
        pages = {arXiv:2406.09246},
          doi = {10.48550/arXiv.2406.09246},
archivePrefix = {arXiv},
       eprint = {2406.09246},
 primaryClass = {cs.RO},
       adsurl = {https://ui.adsabs.harvard.edu/abs/2024arXiv240609246K}
}

@article{hotelling1933analysis,
  title={Analysis of a complex of statistical variables into principal components.},
  author={Hotelling, Harold},
  journal={Journal of educational psychology},
  volume={24},
  number={6},
  pages={417},
  year={1933},
  publisher={Warwick \& York}
}

@article{Baldi1989,
title = {Neural networks and principal component analysis: Learning from examples without local minima},
journal = {Neural Networks},
volume = {2},
number = {1},
pages = {53-58},
year = {1989},
issn = {0893-6080},
doi = {https://doi.org/10.1016/0893-6080(89)90014-2},
url = {https://www.sciencedirect.com/science/article/pii/0893608089900142},
author = {Pierre Baldi and Kurt Hornik}
}

@ARTICLE{Bartlett2022ESR,
       author = {{Bartlett}, Deaglan J. and {Desmond}, Harry and {Ferreira}, Pedro G.},
        title = "{Exhaustive Symbolic Regression}",
      journal = {arXiv e-prints},
         year = 2022,
        month = nov,
          eid = {arXiv:2211.11461},
        pages = {arXiv:2211.11461},
          doi = {10.48550/arXiv.2211.11461},
archivePrefix = {arXiv},
       eprint = {2211.11461},
 primaryClass = {astro-ph.CO},
       adsurl = {https://ui.adsabs.harvard.edu/abs/2022arXiv221111461B}
}

@ARTICLE{Balestriero2025lejepa,
       author = {{Balestriero}, Randall and {LeCun}, Yann},
        title = "{LeJEPA: Provable and Scalable Self-Supervised Learning Without the Heuristics}",
      journal = {arXiv e-prints},
         year = 2025,
        month = nov,
          eid = {arXiv:2511.08544},
        pages = {arXiv:2511.08544},
          doi = {10.48550/arXiv.2511.08544},
archivePrefix = {arXiv},
       eprint = {2511.08544},
 primaryClass = {stat.ML},
       adsurl = {https://ui.adsabs.harvard.edu/abs/2025arXiv251108544B}
}

@ARTICLE{Ivezic2019,
       author = {{Ivezi{\'c}}, {\v{Z}}eljko and {Kahn}, Steven M. and {Tyson}, J. Anthony and {Abel}, Bob and {Acosta}, Emily and {Allsman}, Robyn and {Alonso}, David and {AlSayyad}, Yusra and {Anderson}, Scott F. and {Andrew}, John and {Angel}, James Roger P. and {Angeli}, George Z. and {Ansari}, Reza and {Antilogus}, Pierre and {Araujo}, Constanza and {Armstrong}, Robert and {Arndt}, Kirk T. and {Astier}, Pierre and {Aubourg}, {\'E}ric and {Auza}, Nicole and {Axelrod}, Tim S. and {Bard}, Deborah J. and {Barr}, Jeff D. and {Barrau}, Aurelian and {Bartlett}, James G. and {Bauer}, Amanda E. and {Bauman}, Brian J. and {Baumont}, Sylvain and {Bechtol}, Ellen and {Bechtol}, Keith and {Becker}, Andrew C. and {Becla}, Jacek and {Beldica}, Cristina and {Bellavia}, Steve and {Bianco}, Federica B. and {Biswas}, Rahul and {Blanc}, Guillaume and {Blazek}, Jonathan and {Blandford}, Roger D. and {Bloom}, Josh S. and {Bogart}, Joanne and {Bond}, Tim W. and {Booth}, Michael T. and {Borgland}, Anders W. and {Borne}, Kirk and {Bosch}, James F. and {Boutigny}, Dominique and {Brackett}, Craig A. and {Bradshaw}, Andrew and {Brandt}, William Nielsen and {Brown}, Michael E. and {Bullock}, James S. and {Burchat}, Patricia and {Burke}, David L. and {Cagnoli}, Gianpietro and {Calabrese}, Daniel and {Callahan}, Shawn and {Callen}, Alice L. and {Carlin}, Jeffrey L. and {Carlson}, Erin L. and {Chandrasekharan}, Srinivasan and {Charles-Emerson}, Glenaver and {Chesley}, Steve and {Cheu}, Elliott C. and {Chiang}, Hsin-Fang and {Chiang}, James and {Chirino}, Carol and {Chow}, Derek and {Ciardi}, David R. and {Claver}, Charles F. and {Cohen-Tanugi}, Johann and {Cockrum}, Joseph J. and {Coles}, Rebecca and {Connolly}, Andrew J. and {Cook}, Kem H. and {Cooray}, Asantha and {Covey}, Kevin R. and {Cribbs}, Chris and {Cui}, Wei and {Cutri}, Roc and {Daly}, Philip N. and {Daniel}, Scott F. and {Daruich}, Felipe and {Daubard}, Guillaume and {Daues}, Greg and {Dawson}, William and {Delgado}, Francisco and {Dellapenna}, Alfred and {de Peyster}, Robert and {de Val-Borro}, Miguel and {Digel}, Seth W. and {Doherty}, Peter and {Dubois}, Richard and {Dubois-Felsmann}, Gregory P. and {Durech}, Josef and {Economou}, Frossie and {Eifler}, Tim and {Eracleous}, Michael and {Emmons}, Benjamin L. and {Fausti Neto}, Angelo and {Ferguson}, Henry and {Figueroa}, Enrique and {Fisher-Levine}, Merlin and {Focke}, Warren and {Foss}, Michael D. and {Frank}, James and {Freemon}, Michael D. and {Gangler}, Emmanuel and {Gawiser}, Eric and {Geary}, John C. and {Gee}, Perry and {Geha}, Marla and {Gessner}, Charles J.~B. and {Gibson}, Robert R. and {Gilmore}, D. Kirk and {Glanzman}, Thomas and {Glick}, William and {Goldina}, Tatiana and {Goldstein}, Daniel A. and {Goodenow}, Iain and {Graham}, Melissa L. and {Gressler}, William J. and {Gris}, Philippe and {Guy}, Leanne P. and {Guyonnet}, Augustin and {Haller}, Gunther and {Harris}, Ron and {Hascall}, Patrick A. and {Haupt}, Justine and {Hernandez}, Fabio and {Herrmann}, Sven and {Hileman}, Edward and {Hoblitt}, Joshua and {Hodgson}, John A. and {Hogan}, Craig and {Howard}, James D. and {Huang}, Dajun and {Huffer}, Michael E. and {Ingraham}, Patrick and {Innes}, Walter R. and {Jacoby}, Suzanne H. and {Jain}, Bhuvnesh and {Jammes}, Fabrice and {Jee}, M. James and {Jenness}, Tim and {Jernigan}, Garrett and {Jevremovi{\'c}}, Darko and {Johns}, Kenneth and {Johnson}, Anthony S. and {Johnson}, Margaret W.~G. and {Jones}, R. Lynne and {Juramy-Gilles}, Claire and {Juri{\'c}}, Mario and {Kalirai}, Jason S. and {Kallivayalil}, Nitya J. and {Kalmbach}, Bryce and {Kantor}, Jeffrey P. and {Karst}, Pierre and {Kasliwal}, Mansi M. and {Kelly}, Heather and {Kessler}, Richard and {Kinnison}, Veronica and {Kirkby}, David and {Knox}, Lloyd and {Kotov}, Ivan V. and {Krabbendam}, Victor L. and {Krughoff}, K. Simon and {Kub{\'a}nek}, Petr and {Kuczewski}, John and {Kulkarni}, Shri and {Ku}, John and {Kurita}, Nadine R. and {Lage}, Craig S. and {Lambert}, Ron and {Lange}, Travis and {Langton}, J. Brian and {Le Guillou}, Laurent and {Levine}, Deborah and {Liang}, Ming and {Lim}, Kian-Tat and {Lintott}, Chris J. and {Long}, Kevin E. and {Lopez}, Margaux and {Lotz}, Paul J. and {Lupton}, Robert H. and {Lust}, Nate B. and {MacArthur}, Lauren A. and {Mahabal}, Ashish and {Mandelbaum}, Rachel and {Markiewicz}, Thomas W. and {Marsh}, Darren S. and {Marshall}, Philip J. and {Marshall}, Stuart and {May}, Morgan and {McKercher}, Robert and {McQueen}, Michelle and {Meyers}, Joshua and {Migliore}, Myriam and {Miller}, Michelle and {Mills}, David J.},
        title = "{LSST: From Science Drivers to Reference Design and Anticipated Data Products}",
      journal = {ApJ},
         year = 2019,
        month = mar,
       volume = {873},
       number = {2},
          eid = {111},
        pages = {111},
          doi = {10.3847/1538-4357/ab042c},
archivePrefix = {arXiv},
       eprint = {0805.2366},
 primaryClass = {astro-ph},
       adsurl = {https://ui.adsabs.harvard.edu/abs/2019ApJ...873..111I}
}

@article{Kurucz1993,
  title={SYNTHE spectrum synthesis programs and line data},
  author={Kurucz, Robert L},
  journal={Kurucz CD-Rom},
  year={1993}
}

@ARTICLE{Euclid2025,
       author = {{Euclid Collaboration} and {Mellier}, Y. and {Abdurro'uf} and {Acevedo Barroso}, J.~A. and {Ach{\'u}carro}, A. and {Adamek}, J. and {Adam}, R. and {Addison}, G.~E. and {Aghanim}, N. and {Aguena}, M. and {Ajani}, V. and {Akrami}, Y. and {Al-Bahlawan}, A. and {Alavi}, A. and {Albuquerque}, I.~S. and {Alestas}, G. and {Alguero}, G. and {Allaoui}, A. and {Allen}, S.~W. and {Allevato}, V. and {Alonso-Tetilla}, A.~V. and {Altieri}, B. and {Alvarez-Candal}, A. and {Alvi}, S. and {Amara}, A. and {Amendola}, L. and {Amiaux}, J. and {Andika}, I.~T. and {Andreon}, S. and {Andrews}, A. and {Angora}, G. and {Angulo}, R.~E. and {Annibali}, F. and {Anselmi}, A. and {Anselmi}, S. and {Arcari}, S. and {Archidiacono}, M. and {Aric{\`o}}, G. and {Arnaud}, M. and {Arnouts}, S. and {Asgari}, M. and {Asorey}, J. and {Atayde}, L. and {Atek}, H. and {Atrio-Barandela}, F. and {Aubert}, M. and {Aubourg}, E. and {Auphan}, T. and {Auricchio}, N. and {Aussel}, B. and {Aussel}, H. and {Avelino}, P.~P. and {Avgoustidis}, A. and {Avila}, S. and {Awan}, S. and {Azzollini}, R. and {Baccigalupi}, C. and {Bachelet}, E. and {Bacon}, D. and {Baes}, M. and {Bagley}, M.~B. and {Bahr-Kalus}, B. and {Balaguera-Antolinez}, A. and {Balbinot}, E. and {Balcells}, M. and {Baldi}, M. and {Baldry}, I. and {Balestra}, A. and {Ballardini}, M. and {Ballester}, O. and {Balogh}, M. and {Ba{\~n}ados}, E. and {Barbier}, R. and {Bardelli}, S. and {Baron}, M. and {Barreiro}, T. and {Barrena}, R. and {Barriere}, J.-C. and {Barros}, B.~J. and {Barthelemy}, A. and {Bartolo}, N. and {Basset}, A. and {Battaglia}, P. and {Battisti}, A.~J. and {Baugh}, C.~M. and {Baumont}, L. and {Bazzanini}, L. and {Beaulieu}, J.-P. and {Beckmann}, V. and {Belikov}, A.~N. and {Bel}, J. and {Bellagamba}, F. and {Bella}, M. and {Bellini}, E. and {Benabed}, K. and {Bender}, R. and {Benevento}, G. and {Bennett}, C.~L. and {Benson}, K. and {Bergamini}, P. and {Bermejo-Climent}, J.~R. and {Bernardeau}, F. and {Bertacca}, D. and {Berthe}, M. and {Berthier}, J. and {Bethermin}, M. and {Beutler}, F. and {Bevillon}, C. and {Bhargava}, S. and {Bhatawdekar}, R. and {Bianchi}, D. and {Bisigello}, L. and {Biviano}, A. and {Blake}, R.~P. and {Blanchard}, A. and {Blazek}, J. and {Blot}, L. and {Bosco}, A. and {Bodendorf}, C. and {Boenke}, T. and {B{\"o}hringer}, H. and {Boldrini}, P. and {Bolzonella}, M. and {Bonchi}, A. and {Bonici}, M. and {Bonino}, D. and {Bonino}, L. and {Bonvin}, C. and {Bon}, W. and {Booth}, J.~T. and {Borgani}, S. and {Borlaff}, A.~S. and {Borsato}, E. and {Bose}, B. and {Botticella}, M.~T. and {Boucaud}, A. and {Bouche}, F. and {Boucher}, J.~S. and {Boutigny}, D. and {Bouvard}, T. and {Bouwens}, R. and {Bouy}, H. and {Bowler}, R.~A.~A. and {Bozza}, V. and {Bozzo}, E. and {Branchini}, E. and {Brando}, G. and {Brau-Nogue}, S. and {Brekke}, P. and {Bremer}, M.~N. and {Brescia}, M. and {Breton}, M.-A. and {Brinchmann}, J. and {Brinckmann}, T. and {Brockley-Blatt}, C. and {Brodwin}, M. and {Brouard}, L. and {Brown}, M.~L. and {Bruton}, S. and {Bucko}, J. and {Buddelmeijer}, H. and {Buenadicha}, G. and {Buitrago}, F. and {Burger}, P. and {Burigana}, C. and {Busillo}, V. and {Busonero}, D. and {Cabanac}, R. and {Cabayol-Garcia}, L. and {Cagliari}, M.~S. and {Caillat}, A. and {Caillat}, L. and {Calabrese}, M. and {Calabro}, A. and {Calderone}, G. and {Calura}, F. and {Camacho Quevedo}, B. and {Camera}, S. and {Campos}, L. and {Ca{\~n}as-Herrera}, G. and {Candini}, G.~P. and {Cantiello}, M. and {Capobianco}, V. and {Cappellaro}, E. and {Cappelluti}, N. and {Cappi}, A. and {Caputi}, K.~I. and {Cara}, C. and {Carbone}, C. and {Cardone}, V.~F. and {Carella}, E. and {Carlberg}, R.~G. and {Carle}, M. and {Carminati}, L. and {Caro}, F. and {Carrasco}, J.~M. and {Carretero}, J. and {Carrilho}, P. and {Carron Duque}, J. and {Carry}, B.},
        title = "{Euclid: I. Overview of the Euclid mission}",
      journal = {Astronomy \& Astrophysics},
         year = 2025,
        month = may,
       volume = {697},
          eid = {A1},
        pages = {A1},
          doi = {10.1051/0004-6361/202450810},
archivePrefix = {arXiv},
       eprint = {2405.13491},
 primaryClass = {astro-ph.CO},
       adsurl = {https://ui.adsabs.harvard.edu/abs/2025A&A...697A...1E}
}

@ARTICLE{Ho2024ILI,
       author = {{Ho}, Matthew and {Bartlett}, Deaglan J. and {Chartier}, Nicolas and {Cuesta-Lazaro}, Carolina and {Ding}, Simon and {Lapel}, Axel and {Lemos}, Pablo and {Lovell}, Christopher C. and {Makinen}, T. Lucas and {Modi}, Chirag and {Pandya}, Viraj and {Pandey}, Shivam and {Perez}, Lucia A. and {Wandelt}, Benjamin and {Bryan}, Greg L.},
        title = "{LtU-ILI: An All-in-One Framework for Implicit Inference in Astrophysics and Cosmology}",
      journal = {The Open Journal of Astrophysics},
         year = 2024,
        month = jul,
       volume = {7},
          eid = {54},
        pages = {54},
          doi = {10.33232/001c.120559},
archivePrefix = {arXiv},
       eprint = {2402.05137},
 primaryClass = {astro-ph.IM},
       adsurl = {https://ui.adsabs.harvard.edu/abs/2024OJAp....7E..54H}
}

@ARTICLE{OBriain2021cyclestarnet,
       author = {{O'Briain}, Teaghan and {Ting}, Yuan-Sen and {Fabbro}, S{\'e}bastien and {Yi}, Kwang M. and {Venn}, Kim and {Bialek}, Spencer},
        title = "{Cycle-StarNet: Bridging the Gap between Theory and Data by Leveraging Large Data Sets}",
      journal = {ApJ},
         year = 2021,
        month = jan,
       volume = {906},
       number = {2},
          eid = {130},
        pages = {130},
          doi = {10.3847/1538-4357/abca96},
archivePrefix = {arXiv},
       eprint = {2007.03109},
 primaryClass = {astro-ph.SR},
       adsurl = {https://ui.adsabs.harvard.edu/abs/2021ApJ...906..130O}
}

@ARTICLE{Zhao2026generalize,
       author = {{Zhao}, Xiaosheng and {Ting}, Yuan-Sen and {Wyse}, Rosemary F.~G. and {Szalay}, Alexander S. and {Huang}, Yang and {Dobos}, L{\'a}szl{\'o} and {Budav{\'a}ri}, Tam{\'a}s and {Wei}, Viska},
        title = "{Generalization from Low- to Moderate-Resolution Spectra with Neural Networks for Stellar Parameter Estimation: A Case Study with DESI}",
      journal = {arXiv e-prints},
         year = 2026,
        month = feb,
          eid = {arXiv:2602.15021},
        pages = {arXiv:2602.15021},
          doi = {10.48550/arXiv.2602.15021},
archivePrefix = {arXiv},
       eprint = {2602.15021},
 primaryClass = {astro-ph.SR},
       adsurl = {https://ui.adsabs.harvard.edu/abs/2026arXiv260215021Z}
}

@ARTICLE{Tobias2024XPRVS,
       author = {{Buck}, Tobias and {Schwarz}, Christian},
        title = "{Deep Multimodal Representation Learning for Stellar Spectra}",
      journal = {arXiv e-prints},
         year = 2024,
        month = oct,
          eid = {arXiv:2410.16081},
        pages = {arXiv:2410.16081},
          doi = {10.48550/arXiv.2410.16081},
archivePrefix = {arXiv},
       eprint = {2410.16081},
 primaryClass = {astro-ph.SR},
       adsurl = {https://ui.adsabs.harvard.edu/abs/2024arXiv241016081B}
}

@ARTICLE{bommasani2021foundation,
       author = {{Bommasani}, Rishi and {Hudson}, Drew A. and {Adeli}, Ehsan and {Altman}, Russ and {Arora}, Simran and {von Arx}, Sydney and {Bernstein}, Michael S. and {Bohg}, Jeannette and {Bosselut}, Antoine and {Brunskill}, Emma and {Brynjolfsson}, Erik and {Buch}, Shyamal and {Card}, Dallas and {Castellon}, Rodrigo and {Chatterji}, Niladri and {Chen}, Annie and {Creel}, Kathleen and {Quincy Davis}, Jared and {Demszky}, Dora and {Donahue}, Chris and {Doumbouya}, Moussa and {Durmus}, Esin and {Ermon}, Stefano and {Etchemendy}, John and {Ethayarajh}, Kawin and {Fei-Fei}, Li and {Finn}, Chelsea and {Gale}, Trevor and {Gillespie}, Lauren and {Goel}, Karan and {Goodman}, Noah and {Grossman}, Shelby and {Guha}, Neel and {Hashimoto}, Tatsunori and {Henderson}, Peter and {Hewitt}, John and {Ho}, Daniel E. and {Hong}, Jenny and {Hsu}, Kyle and {Huang}, Jing and {Icard}, Thomas and {Jain}, Saahil and {Jurafsky}, Dan and {Kalluri}, Pratyusha and {Karamcheti}, Siddharth and {Keeling}, Geoff and {Khani}, Fereshte and {Khattab}, Omar and {Koh}, Pang Wei and {Krass}, Mark and {Krishna}, Ranjay and {Kuditipudi}, Rohith and {Kumar}, Ananya and {Ladhak}, Faisal and {Lee}, Mina and {Lee}, Tony and {Leskovec}, Jure and {Levent}, Isabelle and {Li}, Xiang Lisa and {Li}, Xuechen and {Ma}, Tengyu and {Malik}, Ali and {Manning}, Christopher D. and {Mirchandani}, Suvir and {Mitchell}, Eric and {Munyikwa}, Zanele and {Nair}, Suraj and {Narayan}, Avanika and {Narayanan}, Deepak and {Newman}, Ben and {Nie}, Allen and {Niebles}, Juan Carlos and {Nilforoshan}, Hamed and {Nyarko}, Julian and {Ogut}, Giray and {Orr}, Laurel and {Papadimitriou}, Isabel and {Park}, Joon Sung and {Piech}, Chris and {Portelance}, Eva and {Potts}, Christopher and {Raghunathan}, Aditi and {Reich}, Rob and {Ren}, Hongyu and {Rong}, Frieda and {Roohani}, Yusuf and {Ruiz}, Camilo and {Ryan}, Jack and {R{\'e}}, Christopher and {Sadigh}, Dorsa and {Sagawa}, Shiori and {Santhanam}, Keshav and {Shih}, Andy and {Srinivasan}, Krishnan and {Tamkin}, Alex and {Taori}, Rohan and {Thomas}, Armin W. and {Tram{\`e}r}, Florian and {Wang}, Rose E. and {Wang}, William and {Wu}, Bohan and {Wu}, Jiajun and {Wu}, Yuhuai and {Xie}, Sang Michael and {Yasunaga}, Michihiro and {You}, Jiaxuan and {Zaharia}, Matei and {Zhang}, Michael and {Zhang}, Tianyi and {Zhang}, Xikun and {Zhang}, Yuhui and {Zheng}, Lucia and {Zhou}, Kaitlyn and {Liang}, Percy},
        title = "{On the Opportunities and Risks of Foundation Models}",
      journal = {arXiv e-prints},
         year = 2021,
        month = aug,
          eid = {arXiv:2108.07258},
        pages = {arXiv:2108.07258},
          doi = {10.48550/arXiv.2108.07258},
archivePrefix = {arXiv},
       eprint = {2108.07258},
 primaryClass = {cs.LG},
       adsurl = {https://ui.adsabs.harvard.edu/abs/2021arXiv210807258B}
}

@ARTICLE{Pan2024scalinglaw,
       author = {{Pan}, Jia-Shu and {Ting}, Yuan-Sen and {Huang}, Yang and {Yu}, Jie and {Liu}, Ji-Feng},
        title = "{The Scaling Law in Stellar Light Curves}",
      journal = {arXiv e-prints},
         year = 2024,
        month = may,
          eid = {arXiv:2405.17156},
        pages = {arXiv:2405.17156},
          doi = {10.48550/arXiv.2405.17156},
archivePrefix = {arXiv},
       eprint = {2405.17156},
 primaryClass = {astro-ph.IM},
       adsurl = {https://ui.adsabs.harvard.edu/abs/2024arXiv240517156P}
}

@ARTICLE{Saoulis2025transfercosmos,
       author = {{Saoulis}, Alex A. and {Piras}, Davide and {Jeffrey}, Niall and {Mancini}, Alessio Spurio and {Ferreira}, Ana M.~G. and {Joachimi}, Benjamin},
        title = "{Transfer learning for multifidelity simulation-based inference in cosmology}",
      journal = {Monthly Notices of the Royal Astronomical Society},
         year = 2025,
        month = sep,
       volume = {542},
       number = {4},
        pages = {3231-3245},
          doi = {10.1093/mnras/staf1436},
archivePrefix = {arXiv},
       eprint = {2505.21215},
 primaryClass = {astro-ph.CO},
       adsurl = {https://ui.adsabs.harvard.edu/abs/2025MNRAS.542.3231S}
}

@inproceedings{vaswani2017attention,
author = {Vaswani, Ashish and Shazeer, Noam and Parmar, Niki and Uszkoreit, Jakob and Jones, Llion and Gomez, Aidan N. and Kaiser, \L{}ukasz and Polosukhin, Illia},
title = {Attention is all you need},
year = {2017},
isbn = {9781510860964},
publisher = {Curran Associates Inc.},
address = {Red Hook, NY, USA},
booktitle = {Proceedings of the 31st International Conference on Neural Information Processing Systems},
pages = {6000–6010},
numpages = {11},
location = {Long Beach, California, USA},
series = {NIPS'17}
}

@ARTICLE{devlin2018bert,
       author = {{Devlin}, Jacob and {Chang}, Ming-Wei and {Lee}, Kenton and {Toutanova}, Kristina},
        title = "{BERT: Pre-training of Deep Bidirectional Transformers for Language Understanding}",
      journal = {arXiv e-prints},
         year = 2018,
        month = oct,
          eid = {arXiv:1810.04805},
        pages = {arXiv:1810.04805},
          doi = {10.48550/arXiv.1810.04805},
archivePrefix = {arXiv},
       eprint = {1810.04805},
 primaryClass = {cs.CL},
       adsurl = {https://ui.adsabs.harvard.edu/abs/2018arXiv181004805D}
}

@techreport{radford2019gpt2,
  author      = {Radford, Alec and Wu, Jeffrey and Child, Rewon and Luan, David and Amodei, Dario and Sutskever, Ilya},
  title       = {Language Models are Unsupervised Multitask Learners},
  institution = {OpenAI},
  year        = {2019},
  url         = {https://cdn.openai.com/better-language-models/language_models_are_unsupervised_multitask_learners.pdf}
}

@ARTICLE{brown2020gpt3,
       author = {{Brown}, Tom B. and {Mann}, Benjamin and {Ryder}, Nick and {Subbiah}, Melanie and {Kaplan}, Jared and {Dhariwal}, Prafulla and {Neelakantan}, Arvind and {Shyam}, Pranav and {Sastry}, Girish and {Askell}, Amanda and {Agarwal}, Sandhini and {Herbert-Voss}, Ariel and {Krueger}, Gretchen and {Henighan}, Tom and {Child}, Rewon and {Ramesh}, Aditya and {Ziegler}, Daniel M. and {Wu}, Jeffrey and {Winter}, Clemens and {Hesse}, Christopher and {Chen}, Mark and {Sigler}, Eric and {Litwin}, Mateusz and {Gray}, Scott and {Chess}, Benjamin and {Clark}, Jack and {Berner}, Christopher and {McCandlish}, Sam and {Radford}, Alec and {Sutskever}, Ilya and {Amodei}, Dario},
        title = "{Language Models are Few-Shot Learners}",
      journal = {arXiv e-prints},
         year = 2020,
        month = may,
          eid = {arXiv:2005.14165},
        pages = {arXiv:2005.14165},
          doi = {10.48550/arXiv.2005.14165},
archivePrefix = {arXiv},
       eprint = {2005.14165},
 primaryClass = {cs.CL},
       adsurl = {https://ui.adsabs.harvard.edu/abs/2020arXiv200514165B}
}

@ARTICLE{guo2025deepseekr1,
       author = {{Guo}, Daya and {Yang}, Dejian and {Zhang}, Haowei and {Song}, Junxiao and {Wang}, Peiyi and {Zhu}, Qihao and {Xu}, Runxin and {Zhang}, Ruoyu and {Ma}, Shirong and {Bi}, Xiao and {Zhang}, Xiaokang and {Yu}, Xingkai and {Wu}, Yu and {Wu}, Z.~F. and {Gou}, Zhibin and {Shao}, Zhihong and {Li}, Zhuoshu and {Gao}, Ziyi and {Liu}, Aixin and {Xue}, Bing and {Wang}, Bingxuan and {Wu}, Bochao and {Feng}, Bei and {Lu}, Chengda and {Zhao}, Chenggang and {Deng}, Chengqi and {Ruan}, Chong and {Dai}, Damai and {Chen}, Deli and {Ji}, Dongjie and {Li}, Erhang and {Lin}, Fangyun and {Dai}, Fucong and {Luo}, Fuli and {Hao}, Guangbo and {Chen}, Guanting and {Li}, Guowei and {Zhang}, H. and {Xu}, Hanwei and {Ding}, Honghui and {Gao}, Huazuo and {Qu}, Hui and {Li}, Hui and {Guo}, Jianzhong and {Li}, Jiashi and {Chen}, Jingchang and {Yuan}, Jingyang and {Tu}, Jinhao and {Qiu}, Junjie and {Li}, Junlong and {Cai}, J.~L. and {Ni}, Jiaqi and {Liang}, Jian and {Chen}, Jin and {Dong}, Kai and {Hu}, Kai and {You}, Kaichao and {Gao}, Kaige and {Guan}, Kang and {Huang}, Kexin and {Yu}, Kuai and {Wang}, Lean and {Zhang}, Lecong and {Zhao}, Liang and {Wang}, Litong and {Zhang}, Liyue and {Xu}, Lei and {Xia}, Leyi and {Zhang}, Mingchuan and {Zhang}, Minghua and {Tang}, Minghui and {Zhou}, Mingxu and {Li}, Meng and {Wang}, Miaojun and {Li}, Mingming and {Tian}, Ning and {Huang}, Panpan and {Zhang}, Peng and {Wang}, Qiancheng and {Chen}, Qinyu and {Du}, Qiushi and {Ge}, Ruiqi and {Zhang}, Ruisong and {Pan}, Ruizhe and {Wang}, Runji and {Chen}, R.~J. and {Jin}, R.~L. and {Chen}, Ruyi and {Lu}, Shanghao and {Zhou}, Shangyan and {Chen}, Shanhuang and {Ye}, Shengfeng and {Wang}, Shiyu and {Yu}, Shuiping and {Zhou}, Shunfeng and {Pan}, Shuting and {Li}, S.~S. and {Zhou}, Shuang and {Wu}, Shaoqing and {Yun}, Tao and {Pei}, Tian and {Sun}, Tianyu and {Wang}, T. and {Zeng}, Wangding and {Liu}, Wen and {Liang}, Wenfeng and {Gao}, Wenjun and {Yu}, Wenqin and {Zhang}, Wentao and {Xiao}, W.~L. and {An}, Wei and {Liu}, Xiaodong and {Wang}, Xiaohan and {Chen}, Xiaokang and {Nie}, Xiaotao and {Cheng}, Xin and {Liu}, Xin and {Xie}, Xin and {Liu}, Xingchao and {Yang}, Xinyu and {Li}, Xinyuan and {Su}, Xuecheng and {Lin}, Xuheng and {Li}, X.~Q. and {Jin}, Xiangyue and {Shen}, Xiaojin and {Chen}, Xiaosha and {Sun}, Xiaowen and {Wang}, Xiaoxiang and {Song}, Xinnan and {Zhou}, Xinyi and {Wang}, Xianzu and {Shan}, Xinxia and {Li}, Y.~K. and {Wang}, Y.~Q. and {Wei}, Y.~X. and {Zhang}, Yang and {Xu}, Yanhong and {Li}, Yao and {Zhao}, Yao and {Sun}, Yaofeng and {Wang}, Yaohui and {Yu}, Yi and {Zhang}, Yichao and {Shi}, Yifan and {Xiong}, Yiliang and {He}, Ying and {Piao}, Yishi and {Wang}, Yisong and {Tan}, Yixuan and {Ma}, Yiyang and {Liu}, Yiyuan and {Guo}, Yongqiang and {Ou}, Yuan and {Wang}, Yuduan and {Gong}, Yue and {Zou}, Yuheng and {He}, Yujia and {Xiong}, Yunfan and {Luo}, Yuxiang and {You}, Yuxiang and {Liu}, Yuxuan and {Zhou}, Yuyang and {Zhu}, Y.~X. and {Huang}, Yanping and {Li}, Yaohui and {Zheng}, Yi and {Zhu}, Yuchen and {Ma}, Yunxian and {Tang}, Ying and {Zha}, Yukun and {Yan}, Yuting and {Ren}, Z.~Z. and {Ren}, Zehui and {Sha}, Zhangli and {Fu}, Zhe and {Xu}, Zhean and {Xie}, Zhenda and {Zhang}, Zhengyan and {Hao}, Zhewen and {Ma}, Zhicheng and {Yan}, Zhigang and {Wu}, Zhiyu and {Gu}, Zihui and {Zhu}, Zijia and {Liu}, Zijun and {Li}, Zilin and {Xie}, Ziwei and {Song}, Ziyang and {Pan}, Zizheng and {Huang}, Zhen and {Xu}, Zhipeng and {Zhang}, Zhongyu and {Zhang}, Zhen},
        title = "{DeepSeek-R1 incentivizes reasoning in LLMs through reinforcement learning}",
      journal = {Nature},
         year = 2025,
        month = sep,
       volume = {645},
       number = {8081},
        pages = {633-638},
          doi = {10.1038/s41586-025-09422-z},
archivePrefix = {arXiv},
       eprint = {2501.12948},
 primaryClass = {cs.CL},
       adsurl = {https://ui.adsabs.harvard.edu/abs/2025Natur.645..633G}
}

@ARTICLE{hoffmann2022chinchilla,
       author = {{Hoffmann}, Jordan and {Borgeaud}, Sebastian and {Mensch}, Arthur and {Buchatskaya}, Elena and {Cai}, Trevor and {Rutherford}, Eliza and {de Las Casas}, Diego and {Hendricks}, Lisa Anne and {Welbl}, Johannes and {Clark}, Aidan and {Hennigan}, Tom and {Noland}, Eric and {Millican}, Katie and {van den Driessche}, George and {Damoc}, Bogdan and {Guy}, Aurelia and {Osindero}, Simon and {Simonyan}, Karen and {Elsen}, Erich and {Rae}, Jack W. and {Vinyals}, Oriol and {Sifre}, Laurent},
        title = "{Training Compute-Optimal Large Language Models}",
      journal = {arXiv e-prints},
         year = 2022,
        month = mar,
          eid = {arXiv:2203.15556},
        pages = {arXiv:2203.15556},
          doi = {10.48550/arXiv.2203.15556},
archivePrefix = {arXiv},
       eprint = {2203.15556},
 primaryClass = {cs.CL},
       adsurl = {https://ui.adsabs.harvard.edu/abs/2022arXiv220315556H}
}

@ARTICLE{ouyang2022instructgpt,
       author = {{Ouyang}, Long and {Wu}, Jeff and {Jiang}, Xu and {Almeida}, Diogo and {Wainwright}, Carroll L. and {Mishkin}, Pamela and {Zhang}, Chong and {Agarwal}, Sandhini and {Slama}, Katarina and {Ray}, Alex and {Schulman}, John and {Hilton}, Jacob and {Kelton}, Fraser and {Miller}, Luke and {Simens}, Maddie and {Askell}, Amanda and {Welinder}, Peter and {Christiano}, Paul and {Leike}, Jan and {Lowe}, Ryan},
        title = "{Training language models to follow instructions with human feedback}",
      journal = {arXiv e-prints},
         year = 2022,
        month = mar,
          eid = {arXiv:2203.02155},
        pages = {arXiv:2203.02155},
          doi = {10.48550/arXiv.2203.02155},
archivePrefix = {arXiv},
       eprint = {2203.02155},
 primaryClass = {cs.CL},
       adsurl = {https://ui.adsabs.harvard.edu/abs/2022arXiv220302155O}
}

@ARTICLE{rafailov2023dpo,
       author = {{Rafailov}, Rafael and {Sharma}, Archit and {Mitchell}, Eric and {Ermon}, Stefano and {Manning}, Christopher D. and {Finn}, Chelsea},
        title = "{Direct Preference Optimization: Your Language Model is Secretly a Reward Model}",
      journal = {arXiv e-prints},
         year = 2023,
        month = may,
          eid = {arXiv:2305.18290},
        pages = {arXiv:2305.18290},
          doi = {10.48550/arXiv.2305.18290},
archivePrefix = {arXiv},
       eprint = {2305.18290},
 primaryClass = {cs.LG},
       adsurl = {https://ui.adsabs.harvard.edu/abs/2023arXiv230518290R}
}

@ARTICLE{radford2021clip,
       author = {{Radford}, Alec and {Kim}, Jong Wook and {Hallacy}, Chris and {Ramesh}, Aditya and {Goh}, Gabriel and {Agarwal}, Sandhini and {Sastry}, Girish and {Askell}, Amanda and {Mishkin}, Pamela and {Clark}, Jack and {Krueger}, Gretchen and {Sutskever}, Ilya},
        title = "{Learning Transferable Visual Models From Natural Language Supervision}",
      journal = {arXiv e-prints},
         year = 2021,
        month = feb,
          eid = {arXiv:2103.00020},
        pages = {arXiv:2103.00020},
          doi = {10.48550/arXiv.2103.00020},
archivePrefix = {arXiv},
       eprint = {2103.00020},
 primaryClass = {cs.CV},
       adsurl = {https://ui.adsabs.harvard.edu/abs/2021arXiv210300020R}
}

@ARTICLE{rombach2022ldm,
       author = {{Rombach}, Robin and {Blattmann}, Andreas and {Lorenz}, Dominik and {Esser}, Patrick and {Ommer}, Bj{\"o}rn},
        title = "{High-Resolution Image Synthesis with Latent Diffusion Models}",
      journal = {arXiv e-prints},
         year = 2021,
        month = dec,
          eid = {arXiv:2112.10752},
        pages = {arXiv:2112.10752},
          doi = {10.48550/arXiv.2112.10752},
archivePrefix = {arXiv},
       eprint = {2112.10752},
 primaryClass = {cs.CV},
       adsurl = {https://ui.adsabs.harvard.edu/abs/2021arXiv211210752R}
}

@ARTICLE{kirillov2023sam,
       author = {{Kirillov}, Alexander and {Mintun}, Eric and {Ravi}, Nikhila and {Mao}, Hanzi and {Rolland}, Chloe and {Gustafson}, Laura and {Xiao}, Tete and {Whitehead}, Spencer and {Berg}, Alexander C. and {Lo}, Wan-Yen and {Doll{\'a}r}, Piotr and {Girshick}, Ross},
        title = "{Segment Anything}",
      journal = {arXiv e-prints},
         year = 2023,
        month = apr,
          eid = {arXiv:2304.02643},
        pages = {arXiv:2304.02643},
          doi = {10.48550/arXiv.2304.02643},
archivePrefix = {arXiv},
       eprint = {2304.02643},
 primaryClass = {cs.CV},
       adsurl = {https://ui.adsabs.harvard.edu/abs/2023arXiv230402643K}
}

@ARTICLE{assran2023ijepa,
       author = {{Assran}, Mahmoud and {Duval}, Quentin and {Misra}, Ishan and {Bojanowski}, Piotr and {Vincent}, Pascal and {Rabbat}, Michael and {LeCun}, Yann and {Ballas}, Nicolas},
        title = "{Self-Supervised Learning from Images with a Joint-Embedding Predictive Architecture}",
      journal = {arXiv e-prints},
         year = 2023,
        month = jan,
          eid = {arXiv:2301.08243},
        pages = {arXiv:2301.08243},
          doi = {10.48550/arXiv.2301.08243},
archivePrefix = {arXiv},
       eprint = {2301.08243},
 primaryClass = {cs.CV},
       adsurl = {https://ui.adsabs.harvard.edu/abs/2023arXiv230108243A}
}

@ARTICLE{assran2025vjepa2,
       author = {{Assran}, Mido and {Bardes}, Adrien and {Fan}, David and {Garrido}, Quentin and {Howes}, Russell and {Mojtaba} and {Komeili} and {Muckley}, Matthew and {Rizvi}, Ammar and {Roberts}, Claire and {Sinha}, Koustuv and {Zholus}, Artem and {Arnaud}, Sergio and {Gejji}, Abha and {Martin}, Ada and {Hogan}, Francois Robert and {Dugas}, Daniel and {Bojanowski}, Piotr and {Khalidov}, Vasil and {Labatut}, Patrick and {Massa}, Francisco and {Szafraniec}, Marc and {Krishnakumar}, Kapil and {Li}, Yong and {Ma}, Xiaodong and {Chandar}, Sarath and {Meier}, Franziska and {LeCun}, Yann and {Rabbat}, Michael and {Ballas}, Nicolas},
        title = "{V-JEPA 2: Self-Supervised Video Models Enable Understanding, Prediction and Planning}",
      journal = {arXiv e-prints},
         year = 2025,
        month = jun,
          eid = {arXiv:2506.09985},
        pages = {arXiv:2506.09985},
          doi = {10.48550/arXiv.2506.09985},
archivePrefix = {arXiv},
       eprint = {2506.09985},
 primaryClass = {cs.AI},
       adsurl = {https://ui.adsabs.harvard.edu/abs/2025arXiv250609985A}
}

@ARTICLE{parker2024astroclip,
       author = {{Parker}, Liam and {Lanusse}, Francois and {Golkar}, Siavash and {Sarra}, Leopoldo and {Cranmer}, Miles and {Bietti}, Alberto and {Eickenberg}, Michael and {Krawezik}, Geraud and {McCabe}, Michael and {Morel}, Rudy and {Ohana}, Ruben and {Pettee}, Mariel and {R{\'e}galdo-Saint Blancard}, Bruno and {Cho}, Kyunghyun and {Ho}, Shirley and {Polymathic AI Collaboration}},
        title = "{AstroCLIP: a cross-modal foundation model for galaxies}",
      journal = {MNRAS},
         year = 2024,
        month = jul,
       volume = {531},
       number = {4},
        pages = {4990-5011},
          doi = {10.1093/mnras/stae1450},
archivePrefix = {arXiv},
       eprint = {2310.03024},
 primaryClass = {astro-ph.IM},
       adsurl = {https://ui.adsabs.harvard.edu/abs/2024MNRAS.531.4990P}
}

@ARTICLE{donoso2023astromer,
       author = {{Donoso-Oliva}, C. and {Becker}, I. and {Protopapas}, P. and {Cabrera-Vives}, G. and {Vishnu}, M. and {Vardhan}, H.},
        title = "{ASTROMER. A transformer-based embedding for the representation of light curves}",
      journal = {Astronomy and Astrophysics},
         year = 2023,
        month = feb,
       volume = {670},
          eid = {A54},
        pages = {A54},
          doi = {10.1051/0004-6361/202243928},
archivePrefix = {arXiv},
       eprint = {2205.01677},
 primaryClass = {astro-ph.IM},
       adsurl = {https://ui.adsabs.harvard.edu/abs/2023A&A...670A..54D}
}

@ARTICLE{ono2024debiasing,
       author = {{Ono}, Victoria and {Park}, Core Francisco and {Mudur}, Nayantara and {Ni}, Yueying and {Cuesta-Lazaro}, Carolina and {Villaescusa-Navarro}, Francisco},
        title = "{Debiasing with Diffusion: Probabilistic Reconstruction of Dark Matter Fields from Galaxies with CAMELS}",
      journal = {ApJ},
         year = 2024,
        month = aug,
       volume = {970},
       number = {2},
          eid = {174},
        pages = {174},
          doi = {10.3847/1538-4357/ad5957},
archivePrefix = {arXiv},
       eprint = {2403.10648},
 primaryClass = {astro-ph.CO},
       adsurl = {https://ui.adsabs.harvard.edu/abs/2024ApJ...970..174O}
}

@ARTICLE{kaplan2020scaling,
       author = {{Kaplan}, Jared and {McCandlish}, Sam and {Henighan}, Tom and {Brown}, Tom B. and {Chess}, Benjamin and {Child}, Rewon and {Gray}, Scott and {Radford}, Alec and {Wu}, Jeffrey and {Amodei}, Dario},
        title = "{Scaling Laws for Neural Language Models}",
      journal = {arXiv e-prints},
         year = 2020,
        month = jan,
          eid = {arXiv:2001.08361},
        pages = {arXiv:2001.08361},
          doi = {10.48550/arXiv.2001.08361},
archivePrefix = {arXiv},
       eprint = {2001.08361},
 primaryClass = {cs.LG},
       adsurl = {https://ui.adsabs.harvard.edu/abs/2020arXiv200108361K}
}

@ARTICLE{Lastufka2025supervisedcompare,
       author = {{Lastufka}, E. and {Bait}, O. and {Drozdova}, M. and {Kinakh}, V. and {Piras}, D. and {Audard}, M. and {Dessauges-Zavadsky}, M. and {Holotyak}, T. and {Schaerer}, D. and {Voloshynovskiy}, S.},
        title = "{Examining vision foundation models for classification and detection in optical and radio astronomy}",
      journal = {A\&A},
         year = 2025,
        month = nov,
       volume = {703},
          eid = {A217},
        pages = {A217},
          doi = {10.1051/0004-6361/202553691},
archivePrefix = {arXiv},
       eprint = {2409.11175},
 primaryClass = {astro-ph.IM},
       adsurl = {https://ui.adsabs.harvard.edu/abs/2025A&A...703A.217L}
}

@book{bartholomew2011latent,
  title={Latent variable models and factor analysis: A unified approach},
  author={Bartholomew, David J and Knott, Martin and Moustaki, Irini},
  year={2011},
  publisher={John Wiley \& Sons}
}

@book{christopher2006pattern,
  author    = {Bishop, Christopher M.},
  title     = {Pattern Recognition and Machine Learning},
  series    = {Information Science and Statistics},
  publisher = {Springer},
  year      = {2006},
  isbn      = {9780387310732}
}

@ARTICLE{dosovitskiy2021vit,
       author = {{Dosovitskiy}, Alexey and {Beyer}, Lucas and {Kolesnikov}, Alexander and {Weissenborn}, Dirk and {Zhai}, Xiaohua and {Unterthiner}, Thomas and {Dehghani}, Mostafa and {Minderer}, Matthias and {Heigold}, Georg and {Gelly}, Sylvain and {Uszkoreit}, Jakob and {Houlsby}, Neil},
        title = "{An Image is Worth 16x16 Words: Transformers for Image Recognition at Scale}",
      journal = {arXiv e-prints},
         year = 2020,
        month = oct,
          eid = {arXiv:2010.11929},
        pages = {arXiv:2010.11929},
          doi = {10.48550/arXiv.2010.11929},
archivePrefix = {arXiv},
       eprint = {2010.11929},
 primaryClass = {cs.CV},
       adsurl = {https://ui.adsabs.harvard.edu/abs/2020arXiv201011929D}
}

@ARTICLE{gu2023mamba,
       author = {{Gu}, Albert and {Dao}, Tri},
        title = "{Mamba: Linear-Time Sequence Modeling with Selective State Spaces}",
      journal = {arXiv e-prints},
         year = 2023,
        month = dec,
          eid = {arXiv:2312.00752},
        pages = {arXiv:2312.00752},
          doi = {10.48550/arXiv.2312.00752},
archivePrefix = {arXiv},
       eprint = {2312.00752},
 primaryClass = {cs.LG},
       adsurl = {https://ui.adsabs.harvard.edu/abs/2023arXiv231200752G}
}

@ARTICLE{wei2022emergent,
       author = {{Wei}, Jason and {Tay}, Yi and {Bommasani}, Rishi and {Raffel}, Colin and {Zoph}, Barret and {Borgeaud}, Sebastian and {Yogatama}, Dani and {Bosma}, Maarten and {Zhou}, Denny and {Metzler}, Donald and {Chi}, Ed H. and {Hashimoto}, Tatsunori and {Vinyals}, Oriol and {Liang}, Percy and {Dean}, Jeff and {Fedus}, William},
        title = "{Emergent Abilities of Large Language Models}",
      journal = {arXiv e-prints},
         year = 2022,
        month = jun,
          eid = {arXiv:2206.07682},
        pages = {arXiv:2206.07682},
          doi = {10.48550/arXiv.2206.07682},
archivePrefix = {arXiv},
       eprint = {2206.07682},
 primaryClass = {cs.CL},
       adsurl = {https://ui.adsabs.harvard.edu/abs/2022arXiv220607682W}
}

@ARTICLE{gu2022s4,
       author = {{Gu}, Albert and {Goel}, Karan and {R{\'e}}, Christopher},
        title = "{Efficiently Modeling Long Sequences with Structured State Spaces}",
      journal = {arXiv e-prints},
         year = 2021,
        month = oct,
          eid = {arXiv:2111.00396},
        pages = {arXiv:2111.00396},
          doi = {10.48550/arXiv.2111.00396},
archivePrefix = {arXiv},
       eprint = {2111.00396},
 primaryClass = {cs.LG},
       adsurl = {https://ui.adsabs.harvard.edu/abs/2021arXiv211100396G}
}

@ARTICLE{wei2022cot,
       author = {{Wei}, Jason and {Wang}, Xuezhi and {Schuurmans}, Dale and {Bosma}, Maarten and {Ichter}, Brian and {Xia}, Fei and {Chi}, Ed and {Le}, Quoc and {Zhou}, Denny},
        title = "{Chain-of-Thought Prompting Elicits Reasoning in Large Language Models}",
      journal = {arXiv e-prints},
         year = 2022,
        month = jan,
          eid = {arXiv:2201.11903},
        pages = {arXiv:2201.11903},
          doi = {10.48550/arXiv.2201.11903},
archivePrefix = {arXiv},
       eprint = {2201.11903},
 primaryClass = {cs.CL},
       adsurl = {https://ui.adsabs.harvard.edu/abs/2022arXiv220111903W}
}

@ARTICLE{power2022grokking,
       author = {{Power}, Alethea and {Burda}, Yuri and {Edwards}, Harri and {Babuschkin}, Igor and {Misra}, Vedant},
        title = "{Grokking: Generalization Beyond Overfitting on Small Algorithmic Datasets}",
      journal = {arXiv e-prints},
         year = 2022,
        month = jan,
          eid = {arXiv:2201.02177},
        pages = {arXiv:2201.02177},
          doi = {10.48550/arXiv.2201.02177},
archivePrefix = {arXiv},
       eprint = {2201.02177},
 primaryClass = {cs.LG},
       adsurl = {https://ui.adsabs.harvard.edu/abs/2022arXiv220102177P}
}

@ARTICLE{nanda2023progress,
       author = {{Nanda}, Neel and {Chan}, Lawrence and {Lieberum}, Tom and {Smith}, Jess and {Steinhardt}, Jacob},
        title = "{Progress measures for grokking via mechanistic interpretability}",
      journal = {arXiv e-prints},
         year = 2023,
        month = jan,
          eid = {arXiv:2301.05217},
        pages = {arXiv:2301.05217},
          doi = {10.48550/arXiv.2301.05217},
archivePrefix = {arXiv},
       eprint = {2301.05217},
 primaryClass = {cs.LG},
       adsurl = {https://ui.adsabs.harvard.edu/abs/2023arXiv230105217N}
}

@ARTICLE{chameleon2024,
       author = {{Chameleon Team}},
        title = "{Chameleon: Mixed-Modal Early-Fusion Foundation Models}",
      journal = {arXiv e-prints},
         year = 2024,
        month = may,
          eid = {arXiv:2405.09818},
        pages = {arXiv:2405.09818},
          doi = {10.48550/arXiv.2405.09818},
archivePrefix = {arXiv},
       eprint = {2405.09818},
 primaryClass = {cs.CL},
       adsurl = {https://ui.adsabs.harvard.edu/abs/2024arXiv240509818C}
}

@INPROCEEDINGS{Walmsley2022jointschedule,
       author = {{Walmsley}, Mike and {Slijepcevic}, Inigo and {Bowles}, Micah R. and {Scaife}, Anna},
        title = "{Toward Galaxy Foundation Models with Hybrid Contrastive Learning}",
    booktitle = {Machine Learning for Astrophysics},
         year = 2022,
        month = jul,
          eid = {29},
        pages = {29},
          doi = {10.48550/arXiv.2206.11927},
archivePrefix = {arXiv},
       eprint = {2206.11927},
 primaryClass = {cs.CV},
       adsurl = {https://ui.adsabs.harvard.edu/abs/2022mla..confE..29W}
}

@ARTICLE{Schiappa2022videosurvey,
       author = {{Schiappa}, Madeline C. and {Rawat}, Yogesh S. and {Shah}, Mubarak},
        title = "{Self-Supervised Learning for Videos: A Survey}",
      journal = {arXiv e-prints},
         year = 2022,
        month = jun,
          eid = {arXiv:2207.00419},
        pages = {arXiv:2207.00419},
          doi = {10.48550/arXiv.2207.00419},
archivePrefix = {arXiv},
       eprint = {2207.00419},
 primaryClass = {cs.CV},
       adsurl = {https://ui.adsabs.harvard.edu/abs/2022arXiv220700419S}
}

@ARTICLE{hu2021lora,
       author = {{Hu}, Edward J. and {Shen}, Yelong and {Wallis}, Phillip and {Allen-Zhu}, Zeyuan and {Li}, Yuanzhi and {Wang}, Shean and {Wang}, Lu and {Chen}, Weizhu},
        title = "{LoRA: Low-Rank Adaptation of Large Language Models}",
      journal = {arXiv e-prints},
         year = 2021,
        month = jun,
          eid = {arXiv:2106.09685},
        pages = {arXiv:2106.09685},
          doi = {10.48550/arXiv.2106.09685},
archivePrefix = {arXiv},
       eprint = {2106.09685},
 primaryClass = {cs.CL},
       adsurl = {https://ui.adsabs.harvard.edu/abs/2021arXiv210609685H}
}

@ARTICLE{varma2023circuit,
       author = {{Varma}, Vikrant and {Shah}, Rohin and {Kenton}, Zachary and {Kram{\'a}r}, J{\'a}nos and {Kumar}, Ramana},
        title = "{Explaining grokking through circuit efficiency}",
      journal = {arXiv e-prints},
         year = 2023,
        month = sep,
          eid = {arXiv:2309.02390},
        pages = {arXiv:2309.02390},
          doi = {10.48550/arXiv.2309.02390},
archivePrefix = {arXiv},
       eprint = {2309.02390},
 primaryClass = {cs.LG},
       adsurl = {https://ui.adsabs.harvard.edu/abs/2023arXiv230902390V}
}

@ARTICLE{schaeffer2023mirage,
       author = {{Schaeffer}, Rylan and {Miranda}, Brando and {Koyejo}, Sanmi},
        title = "{Are Emergent Abilities of Large Language Models a Mirage?}",
      journal = {arXiv e-prints},
         year = 2023,
        month = apr,
          eid = {arXiv:2304.15004},
        pages = {arXiv:2304.15004},
          doi = {10.48550/arXiv.2304.15004},
archivePrefix = {arXiv},
       eprint = {2304.15004},
 primaryClass = {cs.AI},
       adsurl = {https://ui.adsabs.harvard.edu/abs/2023arXiv230415004S}
}

@ARTICLE{chen2020simclr,
       author = {{Chen}, Ting and {Kornblith}, Simon and {Norouzi}, Mohammad and {Hinton}, Geoffrey},
        title = "{A Simple Framework for Contrastive Learning of Visual Representations}",
      journal = {arXiv e-prints},
         year = 2020,
        month = feb,
          eid = {arXiv:2002.05709},
        pages = {arXiv:2002.05709},
          doi = {10.48550/arXiv.2002.05709},
archivePrefix = {arXiv},
       eprint = {2002.05709},
 primaryClass = {cs.LG},
       adsurl = {https://ui.adsabs.harvard.edu/abs/2020arXiv200205709C}
}

@ARTICLE{he2019Moco,
       author = {{He}, Kaiming and {Fan}, Haoqi and {Wu}, Yuxin and {Xie}, Saining and {Girshick}, Ross},
        title = "{Momentum Contrast for Unsupervised Visual Representation Learning}",
      journal = {arXiv e-prints},
         year = 2019,
        month = nov,
          eid = {arXiv:1911.05722},
        pages = {arXiv:1911.05722},
          doi = {10.48550/arXiv.1911.05722},
archivePrefix = {arXiv},
       eprint = {1911.05722},
 primaryClass = {cs.CV},
       adsurl = {https://ui.adsabs.harvard.edu/abs/2019arXiv191105722H}
}

@ARTICLE{grill2020byol,
       author = {{Grill}, Jean-Bastien and {Strub}, Florian and {Altch{\'e}}, Florent and {Tallec}, Corentin and {Richemond}, Pierre H. and {Buchatskaya}, Elena and {Doersch}, Carl and {Avila Pires}, Bernardo and {Guo}, Zhaohan Daniel and {Gheshlaghi Azar}, Mohammad and {Piot}, Bilal and {Kavukcuoglu}, Koray and {Munos}, R{\'e}mi and {Valko}, Michal},
        title = "{Bootstrap your own latent: A new approach to self-supervised Learning}",
      journal = {arXiv e-prints},
         year = 2020,
        month = jun,
          eid = {arXiv:2006.07733},
        pages = {arXiv:2006.07733},
          doi = {10.48550/arXiv.2006.07733},
archivePrefix = {arXiv},
       eprint = {2006.07733},
 primaryClass = {cs.LG},
       adsurl = {https://ui.adsabs.harvard.edu/abs/2020arXiv200607733G}
}

@ARTICLE{el2024aim,
       author = {{El-Nouby}, Alaaeldin and {Klein}, Michal and {Zhai}, Shuangfei and {Bautista}, Miguel Angel and {Toshev}, Alexander and {Shankar}, Vaishaal and {Susskind}, Joshua M and {Joulin}, Armand},
        title = "{Scalable Pre-training of Large Autoregressive Image Models}",
      journal = {arXiv e-prints},
         year = 2024,
        month = jan,
          eid = {arXiv:2401.08541},
        pages = {arXiv:2401.08541},
          doi = {10.48550/arXiv.2401.08541},
archivePrefix = {arXiv},
       eprint = {2401.08541},
 primaryClass = {cs.CV},
       adsurl = {https://ui.adsabs.harvard.edu/abs/2024arXiv240108541E}
}

@ARTICLE{gilda2020imagequality,
       author = {{Gilda}, Sankalp and {Ting}, Yuan-Sen and {Withington}, Kanoa and {Wilson}, Matthew and {Prunet}, Simon and {Mahoney}, William and {Fabbro}, Sebastien and {Draper}, Stark C. and {Sheinis}, Andrew},
        title = "{Astronomical Image Quality Prediction based on Environmental and Telescope Operating Conditions}",
      journal = {arXiv e-prints},
         year = 2020,
        month = nov,
          eid = {arXiv:2011.03132},
        pages = {arXiv:2011.03132},
          doi = {10.48550/arXiv.2011.03132},
archivePrefix = {arXiv},
       eprint = {2011.03132},
 primaryClass = {astro-ph.IM},
       adsurl = {https://ui.adsabs.harvard.edu/abs/2020arXiv201103132G}
}

@ARTICLE{leung2024stellarfm,
       author = {{Leung}, Henry W. and {Bovy}, Jo},
        title = "{Towards an astronomical foundation model for stars with a transformer-based model}",
      journal = {MNRAS},
         year = 2024,
        month = jan,
       volume = {527},
       number = {1},
        pages = {1494-1520},
          doi = {10.1093/mnras/stad3015},
archivePrefix = {arXiv},
       eprint = {2308.10944},
 primaryClass = {astro-ph.IM},
       adsurl = {https://ui.adsabs.harvard.edu/abs/2024MNRAS.527.1494L}
}

@ARTICLE{parker2025aion,
       author = {{Parker}, Liam and {Lanusse}, Francois and {Shen}, Jeff and {Liu}, Ollie and {Hehir}, Tom and {Sarra}, Leopoldo and {Meyer}, Lucas and {Bowles}, Micah and {Wagner-Carena}, Sebastian and {Qu}, Helen and {Golkar}, Siavash and {Bietti}, Alberto and {Bourfoune}, Hatim and {Casserau}, Nathan and {Cornette}, Pierre and {Hirashima}, Keiya and {Krawezik}, Geraud and {Ohana}, Ruben and {Lourie}, Nicholas and {McCabe}, Michael and {Morel}, Rudy and {Mukhopadhyay}, Payel and {Pettee}, Mariel and {Regaldo-Saint Blancard}, Bruno and {Cho}, Kyunghyun and {Cranmer}, Miles and {Ho}, Shirley},
        title = "{AION-1: Omnimodal Foundation Model for Astronomical Sciences}",
      journal = {arXiv e-prints},
         year = 2025,
        month = oct,
          eid = {arXiv:2510.17960},
        pages = {arXiv:2510.17960},
          doi = {10.48550/arXiv.2510.17960},
archivePrefix = {arXiv},
       eprint = {2510.17960},
 primaryClass = {astro-ph.IM},
       adsurl = {https://ui.adsabs.harvard.edu/abs/2025arXiv251017960P}
}

@ARTICLE{zhao2025specclip,
       author = {{Zhao}, Xiaosheng and {Huang}, Yang and {Xue}, Guirong and {Kong}, Xiao and {Liu}, Jifeng and {Tang}, Xiaoyu and {Beers}, Timothy C. and {Ting}, Yuan-Sen and {Luo}, A.-Li},
        title = "{SpecCLIP: Aligning and Translating Spectroscopic Measurements for Stars}",
      journal = {ApJ},
         year = 2026,
        month = feb,
       volume = {998},
       number = {2},
          eid = {189},
        pages = {189},
          doi = {10.3847/1538-4357/ae2c7e},
archivePrefix = {arXiv},
       eprint = {2507.01939},
 primaryClass = {astro-ph.IM},
       adsurl = {https://ui.adsabs.harvard.edu/abs/2026ApJ...998..189Z}
}

@ARTICLE{cranmer2020sbi,
       author = {{Cranmer}, Kyle and {Brehmer}, Johann and {Louppe}, Gilles},
        title = "{The frontier of simulation-based inference}",
      journal = {Proceedings of the National Academy of Science},
         year = 2020,
        month = dec,
       volume = {117},
       number = {48},
        pages = {30055-30062},
          doi = {10.1073/pnas.1912789117},
archivePrefix = {arXiv},
       eprint = {1911.01429},
 primaryClass = {stat.ML},
       adsurl = {https://ui.adsabs.harvard.edu/abs/2020PNAS..11730055C}
}

@article{mcculloch1943logical,
  title={A logical calculus of the ideas immanent in nervous activity},
  author={McCulloch, Warren S and Pitts, Walter},
  journal={The bulletin of mathematical biophysics},
  volume={5},
  number={4},
  pages={115--133},
  year={1943},
  publisher={Springer}
}

@article{rosenblatt1958perceptron,
  title={The perceptron: a probabilistic model for information storage and organization in the brain.},
  author={Rosenblatt, Frank},
  journal={Psychological review},
  volume={65},
  number={6},
  pages={386},
  year={1958},
  publisher={American Psychological Association}
}

@ARTICLE{lecun1998gradient,
  author={Lecun, Y. and Bottou, L. and Bengio, Y. and Haffner, P.},
  journal={Proceedings of the IEEE}, 
  title={Gradient-based learning applied to document recognition}, 
  year={1998},
  volume={86},
  number={11},
  pages={2278-2324},
  doi={10.1109/5.726791}}

@article{bengio2013representation,
author = {Bengio, Yoshua and Courville, Aaron and Vincent, Pascal},
title = {Representation Learning: A Review and New Perspectives},
year = {2013},
issue_date = {August 2013},
publisher = {IEEE Computer Society},
address = {USA},
volume = {35},
number = {8},
issn = {0162-8828},
url = {https://doi.org/10.1109/TPAMI.2013.50},
doi = {10.1109/TPAMI.2013.50},
journal = {IEEE Transactions on Pattern Analysis and Machine Intelligence},
month = aug,
pages = {1798–1828},
numpages = {31}
}

@article{krizhevsky2012imagenet,
author = {Krizhevsky, Alex and Sutskever, Ilya and Hinton, Geoffrey E.},
title = {ImageNet classification with deep convolutional neural networks},
year = {2017},
issue_date = {June 2017},
publisher = {Association for Computing Machinery},
address = {New York, NY, USA},
volume = {60},
number = {6},
issn = {0001-0782},
url = {https://doi.org/10.1145/3065386},
doi = {10.1145/3065386},
journal = {Commun. ACM},
month = may,
pages = {84–90},
numpages = {7}
}

@ARTICLE{zeiler2014visualizing,
       author = {{Zeiler}, Matthew D and {Fergus}, Rob},
        title = "{Visualizing and Understanding Convolutional Networks}",
      journal = {arXiv e-prints},
         year = 2013,
        month = nov,
          eid = {arXiv:1311.2901},
        pages = {arXiv:1311.2901},
          doi = {10.48550/arXiv.1311.2901},
archivePrefix = {arXiv},
       eprint = {1311.2901},
 primaryClass = {cs.CV},
       adsurl = {https://ui.adsabs.harvard.edu/abs/2013arXiv1311.2901Z}
}

@ARTICLE{cybenko1989approximation,
       author = {{Cybenko}, G.},
        title = "{Approximation by superpositions of a sigmoidal function}",
      journal = {Mathematics of Control, Signals, and Systems},
         year = 1989,
        month = dec,
       volume = {2},
       number = {4},
        pages = {303-314},
          doi = {10.1007/BF02551274},
       adsurl = {https://ui.adsabs.harvard.edu/abs/1989MCSS....2..303C}
}

@article{hornik1991approximation,
title = {Approximation capabilities of multilayer feedforward networks},
journal = {Neural Networks},
volume = {4},
number = {2},
pages = {251-257},
year = {1991},
issn = {0893-6080},
doi = {https://doi.org/10.1016/0893-6080(91)90009-T},
url = {https://www.sciencedirect.com/science/article/pii/089360809190009T},
author = {Kurt Hornik}
}

@article{caruana1997multitask,
  title={Multitask learning},
  author={Caruana, Rich},
  journal={Machine learning},
  volume={28},
  number={1},
  pages={41--75},
  year={1997},
  publisher={Springer}
}

@article{pan2009survey,
author = {Pan, Sinno Jialin and Yang, Qiang},
title = {A Survey on Transfer Learning},
year = {2010},
issue_date = {October 2010},
publisher = {IEEE Educational Activities Department},
address = {USA},
volume = {22},
number = {10},
issn = {1041-4347},
url = {https://doi.org/10.1109/TKDE.2009.191},
doi = {10.1109/TKDE.2009.191},
journal = {IEEE Trans. on Knowl. and Data Eng.},
month = oct,
pages = {1345–1359},
numpages = {15}
}

@ARTICLE{Sharkey2025mechanistic,
       author = {{Sharkey}, Lee and {Chughtai}, Bilal and {Batson}, Joshua and {Lindsey}, Jack and {Wu}, Jeff and {Bushnaq}, Lucius and {Goldowsky-Dill}, Nicholas and {Heimersheim}, Stefan and {Ortega}, Alejandro and {Bloom}, Joseph and {Biderman}, Stella and {Garriga-Alonso}, Adria and {Conmy}, Arthur and {Nanda}, Neel and {Rumbelow}, Jessica and {Wattenberg}, Martin and {Schoots}, Nandi and {Miller}, Joseph and {Michaud}, Eric J. and {Casper}, Stephen and {Tegmark}, Max and {Saunders}, William and {Bau}, David and {Todd}, Eric and {Geiger}, Atticus and {Geva}, Mor and {Hoogland}, Jesse and {Murfet}, Daniel and {McGrath}, Tom},
        title = "{Open Problems in Mechanistic Interpretability}",
      journal = {arXiv e-prints},
         year = 2025,
        month = jan,
          eid = {arXiv:2501.16496},
        pages = {arXiv:2501.16496},
          doi = {10.48550/arXiv.2501.16496},
archivePrefix = {arXiv},
       eprint = {2501.16496},
 primaryClass = {cs.LG},
       adsurl = {https://ui.adsabs.harvard.edu/abs/2025arXiv250116496S}
}

@ARTICLE{he2022mae,
       author = {{He}, Kaiming and {Chen}, Xinlei and {Xie}, Saining and {Li}, Yanghao and {Doll{\'a}r}, Piotr and {Girshick}, Ross},
        title = "{Masked Autoencoders Are Scalable Vision Learners}",
      journal = {arXiv e-prints},
         year = 2021,
        month = nov,
          eid = {arXiv:2111.06377},
        pages = {arXiv:2111.06377},
          doi = {10.48550/arXiv.2111.06377},
archivePrefix = {arXiv},
       eprint = {2111.06377},
 primaryClass = {cs.CV},
       adsurl = {https://ui.adsabs.harvard.edu/abs/2021arXiv211106377H}
}

@ARTICLE{oord2018cpc,
       author = {{van den Oord}, Aaron and {Li}, Yazhe and {Vinyals}, Oriol},
        title = "{Representation Learning with Contrastive Predictive Coding}",
      journal = {arXiv e-prints},
         year = 2018,
        month = jul,
          eid = {arXiv:1807.03748},
        pages = {arXiv:1807.03748},
          doi = {10.48550/arXiv.1807.03748},
archivePrefix = {arXiv},
       eprint = {1807.03748},
 primaryClass = {cs.LG},
       adsurl = {https://ui.adsabs.harvard.edu/abs/2018arXiv180703748V}
}

@misc{lecun2022path,
  author       = {LeCun, Yann},
  title        = {A Path Towards Autonomous Machine Intelligence},
  year         = {2022},
  howpublished = {OpenReview},
  url          = {https://openreview.net/forum?id=BZ5a1r-kVsf}
}

@ARTICLE{battaglia2018relational,
       author = {{Battaglia}, Peter W. and {Hamrick}, Jessica B. and {Bapst}, Victor and {Sanchez-Gonzalez}, Alvaro and {Zambaldi}, Vinicius and {Malinowski}, Mateusz and {Tacchetti}, Andrea and {Raposo}, David and {Santoro}, Adam and {Faulkner}, Ryan and {Gulcehre}, Caglar and {Song}, Francis and {Ballard}, Andrew and {Gilmer}, Justin and {Dahl}, George and {Vaswani}, Ashish and {Allen}, Kelsey and {Nash}, Charles and {Langston}, Victoria and {Dyer}, Chris and {Heess}, Nicolas and {Wierstra}, Daan and {Kohli}, Pushmeet and {Botvinick}, Matt and {Vinyals}, Oriol and {Li}, Yujia and {Pascanu}, Razvan},
        title = "{Relational inductive biases, deep learning, and graph networks}",
      journal = {arXiv e-prints},
         year = 2018,
        month = jun,
          eid = {arXiv:1806.01261},
        pages = {arXiv:1806.01261},
          doi = {10.48550/arXiv.1806.01261},
archivePrefix = {arXiv},
       eprint = {1806.01261},
 primaryClass = {cs.LG},
       adsurl = {https://ui.adsabs.harvard.edu/abs/2018arXiv180601261B}
}

@ARTICLE{bronstein2021geometric,
       author = {{Bronstein}, Michael M. and {Bruna}, Joan and {Cohen}, Taco and {Veli{\v{c}}kovi{\'c}}, Petar},
        title = "{Geometric Deep Learning: Grids, Groups, Graphs, Geodesics, and Gauges}",
      journal = {arXiv e-prints},
         year = 2021,
        month = apr,
          eid = {arXiv:2104.13478},
        pages = {arXiv:2104.13478},
          doi = {10.48550/arXiv.2104.13478},
archivePrefix = {arXiv},
       eprint = {2104.13478},
 primaryClass = {cs.LG},
       adsurl = {https://ui.adsabs.harvard.edu/abs/2021arXiv210413478B}
}

@article{raissi2019pinn,
title = {Physics-informed neural networks: A deep learning framework for solving forward and inverse problems involving nonlinear partial differential equations},
journal = {Journal of Computational Physics},
volume = {378},
pages = {686-707},
year = {2019},
issn = {0021-9991},
doi = {https://doi.org/10.1016/j.jcp.2018.10.045},
url = {https://www.sciencedirect.com/science/article/pii/S0021999118307125},
author = {M. Raissi and P. Perdikaris and G.E. Karniadakis}
}

@article{karniadakis2021physics,
  title={Physics-informed machine learning},
  author={Karniadakis, George Em and Kevrekidis, Ioannis G and Lu, Lu and Perdikaris, Paris and Wang, Sifan and Yang, Liu},
  journal={Nature Reviews Physics},
  volume={3},
  number={6},
  pages={422--440},
  year={2021},
  publisher={Nature Publishing Group UK London}
}

@InProceedings{cohen2016group,
  title = 	 {Group Equivariant Convolutional Networks},
  author = 	 {Cohen, Taco and Welling, Max},
  booktitle = 	 {Proceedings of The 33rd International Conference on Machine Learning},
  pages = 	 {2990--2999},
  year = 	 {2016},
  editor = 	 {Balcan, Maria Florina and Weinberger, Kilian Q.},
  volume = 	 {48},
  series = 	 {Proceedings of Machine Learning Research},
  address = 	 {New York, New York, USA},
  month = 	 {20--22 Jun},
  publisher =    {PMLR},
  url = 	 {https://proceedings.mlr.press/v48/cohenc16.html}
}

@ARTICLE{ting2026deeplearning,
       author = {{Ting}, Yuan-Sen},
        title = "{Deep Learning in Astrophysics}",
      journal = {arXiv e-prints},
         year = 2025,
        month = oct,
          eid = {arXiv:2510.10713},
        pages = {arXiv:2510.10713},
          doi = {10.48550/arXiv.2510.10713},
archivePrefix = {arXiv},
       eprint = {2510.10713},
 primaryClass = {astro-ph.IM},
       adsurl = {https://ui.adsabs.harvard.edu/abs/2025arXiv251010713T}
}

@ARTICLE{alayrac2022flamingo,
       author = {{Alayrac}, Jean-Baptiste and {Donahue}, Jeff and {Luc}, Pauline and {Miech}, Antoine and {Barr}, Iain and {Hasson}, Yana and {Lenc}, Karel and {Mensch}, Arthur and {Millican}, Katie and {Reynolds}, Malcolm and {Ring}, Roman and {Rutherford}, Eliza and {Cabi}, Serkan and {Han}, Tengda and {Gong}, Zhitao and {Samangooei}, Sina and {Monteiro}, Marianne and {Menick}, Jacob and {Borgeaud}, Sebastian and {Brock}, Andrew and {Nematzadeh}, Aida and {Sharifzadeh}, Sahand and {Binkowski}, Mikolaj and {Barreira}, Ricardo and {Vinyals}, Oriol and {Zisserman}, Andrew and {Simonyan}, Karen},
        title = "{Flamingo: a Visual Language Model for Few-Shot Learning}",
      journal = {arXiv e-prints},
         year = 2022,
        month = apr,
          eid = {arXiv:2204.14198},
        pages = {arXiv:2204.14198},
          doi = {10.48550/arXiv.2204.14198},
archivePrefix = {arXiv},
       eprint = {2204.14198},
 primaryClass = {cs.CV},
       adsurl = {https://ui.adsabs.harvard.edu/abs/2022arXiv220414198A}
}

@ARTICLE{liang2022modalitygap,
       author = {{Liang}, Weixin and {Zhang}, Yuhui and {Kwon}, Yongchan and {Yeung}, Serena and {Zou}, James},
        title = "{Mind the Gap: Understanding the Modality Gap in Multi-modal Contrastive Representation Learning}",
      journal = {arXiv e-prints},
         year = 2022,
        month = mar,
          eid = {arXiv:2203.02053},
        pages = {arXiv:2203.02053},
          doi = {10.48550/arXiv.2203.02053},
archivePrefix = {arXiv},
       eprint = {2203.02053},
 primaryClass = {cs.CL},
       adsurl = {https://ui.adsabs.harvard.edu/abs/2022arXiv220302053L}
}

@ARTICLE{yuan2021multimodal,
       author = {{Yuan}, Xin and {Lin}, Zhe and {Kuen}, Jason and {Zhang}, Jianming and {Wang}, Yilin and {Maire}, Michael and {Kale}, Ajinkya and {Faieta}, Baldo},
        title = "{Multimodal Contrastive Training for Visual Representation Learning}",
      journal = {arXiv e-prints},
         year = 2021,
        month = apr,
          eid = {arXiv:2104.12836},
        pages = {arXiv:2104.12836},
          doi = {10.48550/arXiv.2104.12836},
archivePrefix = {arXiv},
       eprint = {2104.12836},
 primaryClass = {cs.CV},
       adsurl = {https://ui.adsabs.harvard.edu/abs/2021arXiv210412836Y}
}

@article{fukushima1980neocognitron,
  title={Neocognitron: A self-organizing neural network model for a mechanism of pattern recognition unaffected by shift in position},
  author={Fukushima, Kunihiko},
  journal={Biological cybernetics},
  volume={36},
  number={4},
  pages={193--202},
  year={1980},
  publisher={Springer}
}

@article{rumelhart1986learning,
  title={Learning representations by back-propagating errors},
  author={Rumelhart, David E and Hinton, Geoffrey E and Williams, Ronald J},
  journal={nature},
  volume={323},
  number={6088},
  pages={533--536},
  year={1986},
  publisher={Nature Publishing Group UK London}
}

@article{erhan2010why,
  author  = {Dumitru Erhan and Yoshua Bengio and Aaron Courville and Pierre-Antoine Manzagol and Pascal Vincent and Samy Bengio},
  title   = {Why Does Unsupervised Pre-training Help Deep Learning?},
  journal = {Journal of Machine Learning Research},
  year    = {2010},
  volume  = {11},
  number  = {19},
  pages   = {625--660},
  url     = {http://jmlr.org/papers/v11/erhan10a.html}
}

@inproceedings{yosinski2014transferable,
author = {Yosinski, Jason and Clune, Jeff and Bengio, Yoshua and Lipson, Hod},
title = {How transferable are features in deep neural networks?},
year = {2014},
publisher = {MIT Press},
address = {Cambridge, MA, USA},
booktitle = {Proceedings of the 28th International Conference on Neural Information Processing Systems - Volume 2},
pages = {3320–3328},
numpages = {9},
location = {Montreal, Canada},
series = {NIPS'14}
}

@ARTICLE{kornblith2019better,
       author = {{Kornblith}, Simon and {Shlens}, Jonathon and {Le}, Quoc V.},
        title = "{Do Better ImageNet Models Transfer Better?}",
      journal = {arXiv e-prints},
         year = 2018,
        month = may,
          eid = {arXiv:1805.08974},
        pages = {arXiv:1805.08974},
          doi = {10.48550/arXiv.1805.08974},
archivePrefix = {arXiv},
       eprint = {1805.08974},
 primaryClass = {cs.CV},
       adsurl = {https://ui.adsabs.harvard.edu/abs/2018arXiv180508974K}
}

@ARTICLE{zhang2024maven,
       author = {{Zhang}, Gemma and {Helfer}, Thomas and {Gagliano}, Alexander T. and {Mishra-Sharma}, Siddharth and {Ashley Villar}, V.},
        title = "{Maven: a multimodal foundation model for supernova science}",
      journal = {Machine Learning: Science and Technology},
         year = 2024,
        month = dec,
       volume = {5},
       number = {4},
          eid = {045069},
        pages = {045069},
          doi = {10.1088/2632-2153/ad990d},
archivePrefix = {arXiv},
       eprint = {2408.16829},
 primaryClass = {astro-ph.HE},
       adsurl = {https://ui.adsabs.harvard.edu/abs/2024MLS&T...5d5069Z}
}

@ARTICLE{rizhko2024astrom3,
       author = {{Rizhko}, M. and {Bloom}, J.~S.},
        title = "{AstroM$^{3}$: A Self-supervised Multimodal Model for Astronomy}",
      journal = {AJ},
         year = 2025,
        month = jul,
       volume = {170},
       number = {1},
          eid = {28},
        pages = {28},
          doi = {10.3847/1538-3881/adcbad},
archivePrefix = {arXiv},
       eprint = {2411.08842},
 primaryClass = {astro-ph.IM},
       adsurl = {https://ui.adsabs.harvard.edu/abs/2025AJ....170...28R}
}

@ARTICLE{zuo2025falco,
       author = {{Zuo}, Xiaoxiong and {Tao}, Yihan and {Huang}, Yang and {Kang}, Zhixuan and {Chen}, Huaxi and {Cui}, Chenzhou and {Pan}, Jiashu and {Kong}, Xiao and {Ting}, Yuan-Sen and {Tang}, Xiaoyu and {Han}, Henggeng and {Mu}, Haiyang and {Xu}, Yunfei and {Fan}, Dongwei and {Xue}, Guirong and {Luo}, Ali and {Liu}, Jifeng},
        title = "{FALCO: Foundation Model of Astronomical Light Curves for Time Domain Astronomy. Implementation and Applications on Kepler Data}",
      journal = {AJ},
         year = 2026,
        month = jan,
       volume = {171},
       number = {1},
          eid = {10},
        pages = {10},
          doi = {10.3847/1538-3881/ae1467},
archivePrefix = {arXiv},
       eprint = {2504.20290},
 primaryClass = {astro-ph.IM},
       adsurl = {https://ui.adsabs.harvard.edu/abs/2026AJ....171...10Z}
}

@ARTICLE{koblischke2024spectrafm,
       author = {{Koblischke}, Nolan and {Bovy}, Jo},
        title = "{SpectraFM: Tuning into Stellar Foundation Models}",
      journal = {arXiv e-prints},
         year = 2024,
        month = nov,
          eid = {arXiv:2411.04750},
        pages = {arXiv:2411.04750},
          doi = {10.48550/arXiv.2411.04750},
archivePrefix = {arXiv},
       eprint = {2411.04750},
 primaryClass = {astro-ph.IM},
       adsurl = {https://ui.adsabs.harvard.edu/abs/2024arXiv241104750K}
}

@ARTICLE{rozanski2025scaling,
       author = {{R{\'o}{\.z}a{\'n}ski}, Tomasz and {Ting}, Yuan-Sen},
        title = "{Scaling Laws for Emulation of Stellar Spectra}",
      journal = {The Open Journal of Astrophysics},
         year = 2025,
        month = jun,
       volume = {8},
          eid = {69},
        pages = {69},
          doi = {10.33232/001c.140607},
archivePrefix = {arXiv},
       eprint = {2503.18617},
 primaryClass = {astro-ph.IM},
       adsurl = {https://ui.adsabs.harvard.edu/abs/2025OJAp....8E..69R}
}

@ARTICLE{zhao2025lora,
       author = {{Zhao}, Xiaosheng and {Ting}, Yuan-Sen and {Szalay}, Alexander S. and {Huang}, Yang},
        title = "{Finetuning Stellar Spectra Foundation Models with LoRA}",
      journal = {arXiv e-prints},
         year = 2025,
        month = jul,
          eid = {arXiv:2507.20972},
        pages = {arXiv:2507.20972},
          doi = {10.48550/arXiv.2507.20972},
archivePrefix = {arXiv},
       eprint = {2507.20972},
 primaryClass = {astro-ph.IM},
       adsurl = {https://ui.adsabs.harvard.edu/abs/2025arXiv250720972Z}
}

@ARTICLE{smith2024astropt,
       author = {{Smith}, Michael J. and {Roberts}, Ryan J. and {Angeloudi}, Eirini and {Huertas-Company}, Marc},
        title = "{AstroPT: Scaling Large Observation Models for Astronomy}",
      journal = {arXiv e-prints},
         year = 2024,
        month = may,
          eid = {arXiv:2405.14930},
        pages = {arXiv:2405.14930},
          doi = {10.48550/arXiv.2405.14930},
archivePrefix = {arXiv},
       eprint = {2405.14930},
 primaryClass = {astro-ph.IM},
       adsurl = {https://ui.adsabs.harvard.edu/abs/2024arXiv240514930S}
}

@ARTICLE{Kendall2017multi,
       author = {{Kendall}, Alex and {Gal}, Yarin and {Cipolla}, Roberto},
        title = "{Multi-Task Learning Using Uncertainty to Weigh Losses for Scene Geometry and Semantics}",
      journal = {arXiv e-prints},
         year = 2017,
        month = may,
          eid = {arXiv:1705.07115},
        pages = {arXiv:1705.07115},
          doi = {10.48550/arXiv.1705.07115},
archivePrefix = {arXiv},
       eprint = {1705.07115},
 primaryClass = {cs.CV},
       adsurl = {https://ui.adsabs.harvard.edu/abs/2017arXiv170507115K}
}

@ARTICLE{Sener2018multi,
       author = {{Sener}, Ozan and {Koltun}, Vladlen},
        title = "{Multi-Task Learning as Multi-Objective Optimization}",
      journal = {arXiv e-prints},
         year = 2018,
        month = oct,
          eid = {arXiv:1810.04650},
        pages = {arXiv:1810.04650},
          doi = {10.48550/arXiv.1810.04650},
archivePrefix = {arXiv},
       eprint = {1810.04650},
 primaryClass = {cs.LG},
       adsurl = {https://ui.adsabs.harvard.edu/abs/2018arXiv181004650S}
}

@ARTICLE{Masters2019galaxyzoo,
       author = {{Masters}, Karen L. and {Lintott}, Chris J. and {Hart}, Ross E. and {Kruk}, Sandor J. and {Smethurst}, Rebecca J. and {Casteels}, Kevin V. and {Keel}, William C. and {Simmons}, Brooke D. and {Stanescu}, Dennis O. and {Tate}, Jean and {Tomi}, Satoshi},
        title = "{Galaxy Zoo: unwinding the winding problem - observations of spiral bulge prominence and arm pitch angles suggest local spiral galaxies are winding}",
      journal = {MNRAS},
         year = 2019,
        month = aug,
       volume = {487},
       number = {2},
        pages = {1808-1820},
          doi = {10.1093/mnras/stz1153},
archivePrefix = {arXiv},
       eprint = {1904.11436},
 primaryClass = {astro-ph.GA},
       adsurl = {https://ui.adsabs.harvard.edu/abs/2019MNRAS.487.1808M}
}

@ARTICLE{hayat2021ssl,
       author = {{Hayat}, Md Abul and {Stein}, George and {Harrington}, Peter and {Luki{\'c}}, Zarija and {Mustafa}, Mustafa},
        title = "{Self-supervised Representation Learning for Astronomical Images}",
      journal = {ApJL},
         year = 2021,
        month = apr,
       volume = {911},
       number = {2},
          eid = {L33},
        pages = {L33},
          doi = {10.3847/2041-8213/abf2c7},
archivePrefix = {arXiv},
       eprint = {2012.13083},
 primaryClass = {astro-ph.IM},
       adsurl = {https://ui.adsabs.harvard.edu/abs/2021ApJ...911L..33H}
}

@ARTICLE{stein2022lenses,
       author = {{Stein}, George and {Blaum}, Jacqueline and {Harrington}, Peter and {Medan}, Tomislav and {Luki{\'c}}, Zarija},
        title = "{Mining for Strong Gravitational Lenses with Self-supervised Learning}",
      journal = {ApJ},
         year = 2022,
        month = jun,
       volume = {932},
       number = {2},
          eid = {107},
        pages = {107},
          doi = {10.3847/1538-4357/ac6d63},
archivePrefix = {arXiv},
       eprint = {2110.00023},
 primaryClass = {astro-ph.IM},
       adsurl = {https://ui.adsabs.harvard.edu/abs/2022ApJ...932..107S}
}

@ARTICLE{shen2025universal,
       author = {{Shen}, Jeff and {Lanusse}, Francois and {Holden Parker}, Liam and {Liu}, Ollie and {Hehir}, Tom and {Sarra}, Leopoldo and {Meyer}, Lucas and {Bowles}, Micah and {Wagner-Carena}, Sebastian and {Wagner-Carena}, Sebastian and {Qu}, Helen and {Golkar}, Siavash and {Bietti}, Alberto and {Bourfoune}, Hatim and {Cassereau}, Nathan and {Cornette}, Pierre and {Hirashima}, Keiya and {Krawezik}, Geraud and {Ohana}, Ruben and {Lourie}, Nicholas and {McCabe}, Michael and {Morel}, Rudy and {Mukhopadhyay}, Payel and {Pettee}, Mariel and {R{\'e}galdo-Saint Blancard}, Bruno and {Cho}, Kyunghyun and {Cranmer}, Miles and {Ho}, Shirley},
        title = "{Universal Spectral Tokenization via Self-Supervised Panchromatic Representation Learning}",
      journal = {arXiv e-prints},
         year = 2025,
        month = oct,
          eid = {arXiv:2510.17959},
        pages = {arXiv:2510.17959},
          doi = {10.48550/arXiv.2510.17959},
archivePrefix = {arXiv},
       eprint = {2510.17959},
 primaryClass = {astro-ph.IM},
       adsurl = {https://ui.adsabs.harvard.edu/abs/2025arXiv251017959S}
}

@ARTICLE{ore2024skatr,
       author = {{Ore}, Ayodele and {Heneka}, Caroline and {Plehn}, Tilman},
        title = "{SKATR: A self-supervised summary transformer for SKA}",
      journal = {SciPost Physics},
         year = 2025,
        month = may,
       volume = {18},
       number = {5},
          eid = {155},
        pages = {155},
          doi = {10.21468/SciPostPhys.18.5.155},
archivePrefix = {arXiv},
       eprint = {2410.18899},
 primaryClass = {astro-ph.IM},
       adsurl = {https://ui.adsabs.harvard.edu/abs/2025ScPP...18..155O}
}

@ARTICLE{Kamai2025dual,
       author = {{Kamai}, Ilay and {Bronstein}, Alex M. and {Perets}, Hagai B.},
        title = "{Machine Learning Inference of Stellar Properties Using Integrated Photometric and Spectroscopic Data}",
      journal = {ApJ},
         year = 2025,
        month = nov,
       volume = {994},
       number = {1},
          eid = {110},
        pages = {110},
          doi = {10.3847/1538-4357/ae0cbc},
archivePrefix = {arXiv},
       eprint = {2507.10666},
 primaryClass = {astro-ph.SR},
       adsurl = {https://ui.adsabs.harvard.edu/abs/2025ApJ...994..110K}
}

@ARTICLE{mikolov2013word2vec,
       author = {{Mikolov}, Tomas and {Chen}, Kai and {Corrado}, Greg and {Dean}, Jeffrey},
        title = "{Efficient Estimation of Word Representations in Vector Space}",
      journal = {arXiv e-prints},
         year = 2013,
        month = jan,
          eid = {arXiv:1301.3781},
        pages = {arXiv:1301.3781},
          doi = {10.48550/arXiv.1301.3781},
archivePrefix = {arXiv},
       eprint = {1301.3781},
 primaryClass = {cs.CL},
       adsurl = {https://ui.adsabs.harvard.edu/abs/2013arXiv1301.3781M}
}

@article{hochreiter1997lstm,
author = {Hochreiter, Sepp and Schmidhuber, J\"{u}rgen},
title = {Long Short-Term Memory},
year = {1997},
issue_date = {November 15, 1997},
publisher = {MIT Press},
address = {Cambridge, MA, USA},
volume = {9},
number = {8},
issn = {0899-7667},
url = {https://doi.org/10.1162/neco.1997.9.8.1735},
doi = {10.1162/neco.1997.9.8.1735},
journal = {Neural Comput.},
month = nov,
pages = {1735–1780},
numpages = {46}
}

@inproceedings{sutskever2014seq2seq,
author = {Sutskever, Ilya and Vinyals, Oriol and Le, Quoc V.},
title = {Sequence to sequence learning with neural networks},
year = {2014},
publisher = {MIT Press},
address = {Cambridge, MA, USA},
booktitle = {Proceedings of the 28th International Conference on Neural Information Processing Systems - Volume 2},
pages = {3104–3112},
numpages = {9},
location = {Montreal, Canada},
series = {NIPS'14}
}

@ARTICLE{bahdanau2015attention,
       author = {{Bahdanau}, Dzmitry and {Cho}, Kyunghyun and {Bengio}, Yoshua},
        title = "{Neural Machine Translation by Jointly Learning to Align and Translate}",
      journal = {arXiv e-prints},
         year = 2014,
        month = sep,
          eid = {arXiv:1409.0473},
        pages = {arXiv:1409.0473},
          doi = {10.48550/arXiv.1409.0473},
archivePrefix = {arXiv},
       eprint = {1409.0473},
 primaryClass = {cs.CL},
       adsurl = {https://ui.adsabs.harvard.edu/abs/2014arXiv1409.0473B}
}

@ARTICLE{peters2018elmo,
       author = {{Peters}, Matthew E. and {Neumann}, Mark and {Iyyer}, Mohit and {Gardner}, Matt and {Clark}, Christopher and {Lee}, Kenton and {Zettlemoyer}, Luke},
        title = "{Deep contextualized word representations}",
      journal = {arXiv e-prints},
         year = 2018,
        month = feb,
          eid = {arXiv:1802.05365},
        pages = {arXiv:1802.05365},
          doi = {10.48550/arXiv.1802.05365},
archivePrefix = {arXiv},
       eprint = {1802.05365},
 primaryClass = {cs.CL},
       adsurl = {https://ui.adsabs.harvard.edu/abs/2018arXiv180205365P}
}

@ARTICLE{wang2024llmsurvey,
       author = {{Wang}, Zichong and {Chu}, Zhibo and {Viet Doan}, Thang and {Ni}, Shiwen and {Yang}, Min and {Zhang}, Wenbin},
        title = "{History, Development, and Principles of Large Language Models-An Introductory Survey}",
      journal = {arXiv e-prints},
         year = 2024,
        month = feb,
          eid = {arXiv:2402.06853},
        pages = {arXiv:2402.06853},
          doi = {10.48550/arXiv.2402.06853},
archivePrefix = {arXiv},
       eprint = {2402.06853},
 primaryClass = {cs.CL},
       adsurl = {https://ui.adsabs.harvard.edu/abs/2024arXiv240206853W}
}

@ARTICLE{caron2021dino,
       author = {{Caron}, Mathilde and {Touvron}, Hugo and {Misra}, Ishan and {J{\'e}gou}, Herv{\'e} and {Mairal}, Julien and {Bojanowski}, Piotr and {Joulin}, Armand},
        title = "{Emerging Properties in Self-Supervised Vision Transformers}",
      journal = {arXiv e-prints},
         year = 2021,
        month = apr,
          eid = {arXiv:2104.14294},
        pages = {arXiv:2104.14294},
          doi = {10.48550/arXiv.2104.14294},
archivePrefix = {arXiv},
       eprint = {2104.14294},
 primaryClass = {cs.CV},
       adsurl = {https://ui.adsabs.harvard.edu/abs/2021arXiv210414294C}
}

@ARTICLE{oquab2023dinov2,
       author = {{Oquab}, Maxime and {Darcet}, Timoth{\'e}e and {Moutakanni}, Th{\'e}o and {Vo}, Huy and {Szafraniec}, Marc and {Khalidov}, Vasil and {Fernandez}, Pierre and {Haziza}, Daniel and {Massa}, Francisco and {El-Nouby}, Alaaeldin and {Assran}, Mahmoud and {Ballas}, Nicolas and {Galuba}, Wojciech and {Howes}, Russell and {Huang}, Po-Yao and {Li}, Shang-Wen and {Misra}, Ishan and {Rabbat}, Michael and {Sharma}, Vasu and {Synnaeve}, Gabriel and {Xu}, Hu and {Jegou}, Herv{\'e} and {Mairal}, Julien and {Labatut}, Patrick and {Joulin}, Armand and {Bojanowski}, Piotr},
        title = "{DINOv2: Learning Robust Visual Features without Supervision}",
      journal = {arXiv e-prints},
         year = 2023,
        month = apr,
          eid = {arXiv:2304.07193},
        pages = {arXiv:2304.07193},
          doi = {10.48550/arXiv.2304.07193},
archivePrefix = {arXiv},
       eprint = {2304.07193},
 primaryClass = {cs.CV},
       adsurl = {https://ui.adsabs.harvard.edu/abs/2023arXiv230407193O}
}

@ARTICLE{awais2024visionfm,
       author = {{Awais}, Muhammad and {Naseer}, Muzammal and {Khan}, Salman and {Anwer}, Rao Muhammad and {Cholakkal}, Hisham and {Shah}, Mubarak and {Yang}, Ming-Hsuan and {Shahbaz Khan}, Fahad},
        title = "{Foundational Models Defining a New Era in Vision: A Survey and Outlook}",
      journal = {arXiv e-prints},
         year = 2023,
        month = jul,
          eid = {arXiv:2307.13721},
        pages = {arXiv:2307.13721},
          doi = {10.48550/arXiv.2307.13721},
archivePrefix = {arXiv},
       eprint = {2307.13721},
 primaryClass = {cs.CV},
       adsurl = {https://ui.adsabs.harvard.edu/abs/2023arXiv230713721A}
}

@ARTICLE{greydanus2019hamiltonian,
       author = {{Greydanus}, Sam and {Dzamba}, Misko and {Yosinski}, Jason},
        title = "{Hamiltonian Neural Networks}",
      journal = {arXiv e-prints},
         year = 2019,
        month = jun,
          eid = {arXiv:1906.01563},
        pages = {arXiv:1906.01563},
          doi = {10.48550/arXiv.1906.01563},
archivePrefix = {arXiv},
       eprint = {1906.01563},
 primaryClass = {cs.NE},
       adsurl = {https://ui.adsabs.harvard.edu/abs/2019arXiv190601563G}
}

@ARTICLE{ho2022multifidelity,
       author = {{Ho}, Ming-Feng and {Bird}, Simeon and {Shelton}, Christian R.},
        title = "{Multifidelity emulation for the matter power spectrum using Gaussian processes}",
      journal = {MNRAS},
         year = 2022,
        month = jan,
       volume = {509},
       number = {2},
        pages = {2551-2565},
          doi = {10.1093/mnras/stab3114},
archivePrefix = {arXiv},
       eprint = {2105.01081},
 primaryClass = {astro-ph.CO},
       adsurl = {https://ui.adsabs.harvard.edu/abs/2022MNRAS.509.2551H}
}

@book{jelinek1997statistical,
  title = {Statistical Methods for Speech Recognition},
  author = {Jelinek, Frederick},
  publisher = {MIT Press}, address = {Cambridge, MA}, year = {1997}
}

@inproceedings{mikolov2010recurrent,
  title={Recurrent neural network based language model.},
  author={Mikolov, Tomas and Karafi{\'a}t, Martin and Burget, Lukas and Cernock{\`y}, Jan and Khudanpur, Sanjeev},
  booktitle={Interspeech},
  volume={2},
  number={3},
  pages={1045--1048},
  year={2010},
  organization={Makuhari}
}

@ARTICLE{lavie2024inductivebias,
       author = {{Lavie}, Itay and {Gur-Ari}, Guy and {Ringel}, Zohar},
        title = "{Towards Understanding Inductive Bias in Transformers: A View From Infinity}",
      journal = {arXiv e-prints},
         year = 2024,
        month = feb,
          eid = {arXiv:2402.05173},
        pages = {arXiv:2402.05173},
          doi = {10.48550/arXiv.2402.05173},
archivePrefix = {arXiv},
       eprint = {2402.05173},
 primaryClass = {cs.LG},
       adsurl = {https://ui.adsabs.harvard.edu/abs/2024arXiv240205173L}
}

\end{document}